\documentclass[%
 aapm,
 mph,%
 amsmath,amssymb,
preprint
]{revtex4-2}

\usepackage{graphicx}
\usepackage{dcolumn}
\usepackage{bm}
\usepackage[final]{changes}

\definecolor{myblue}{rgb}{0.0, 0.2, 0.8} 
\colorlet{Changes@Added}{myblue}
\colorlet{Changes@Deleted}{myblue}
\colorlet{Changes@Replaced}{myblue}
\setauthormarkup{} 
\usepackage{amsmath}
\usepackage{geometry}
\usepackage{placeins}  
\usepackage{multirow}
\usepackage{tikz}

\begin{document}

\preprint{AAPM/123-QED}

\title[]{Effective conductivity of conduit networks with random conductivities}

\author{Iv\'an Colecchio}
\author{Elora Le Gall}%
\author{Beno{\^i}t Noetinger}
  \email{benoit.noetinger@ifpen.fr}
\affiliation{ 
IFP Energies Nouvelles  1 \& 4, avenue de Bois-Pr{\'e}au, 92852 Rueil-Malmaison Cedex - France
}%

\date{\today}

\begin{abstract}
The effective conductivity ($T^{eff}$) of 2D and 3D Random Resistor Networks (RRNs) with random edge conductivity \replaced{is}{are} studied. The combined influence of geometrical disorder, which controls the overall connectivity of the medium\added{,} and leads to percolation effects, and conductivity randomness is investigated. A formula incorporating connectivity aspects and second-order averaging methods, widely used in the stochastic hydrology community, is derived and extrapolated to higher orders using a power averaging formula based on a mean-field argument. This approach highlights the role of the so-called resistance distance\added{,} introduced by graph theorists. Simulations are performed on various RRN geometries constructed from 2D and 3D bond-percolation lattices. The results confirm the robustness of the power averaging technique and the relevance of the mean-field assumption.

\end{abstract}

\keywords{Random network, random conductivity, averaging, perturbation theory, mean-field, power-averaging, numerical tests}
\maketitle

\section{\label{sec:Intro}Introduction}
Modeling transport in strongly heterogeneous materials is a generic issue that was addressed long ago in the 19th century\cite{maxwell1873treatise}, and later \replaced{enriched by others}{enriched} \added{such as} \cite{landau1960lifshitz, hashin1962variational, matheron1967, torquato2002random} \added{in their respective research areas}. Applications range from modeling mechanical properties of composite media, heat \replaced{transfer}{transfers}, electrical currents, to momentum transfer in porous media at different scales. Restricting ourselves mainly to the former area, many theoretical approaches have been proposed by theoretical hydrologists \citep{gelhar1993stochastic, dagan1989flow, indelman1994higher, abramovich1995effective, renard1997calculating}, by mathematicians \citep{jikov2012homogenization, armstrong2019quantitative}\added{,} and \added{by} physicists \citep{king1987use, quintard1994transport, n1994effective, noetinger1998use, hristopulos1999renormalization, NOETINGER2000353, teodorovich2002renormalization, attinger2003generalized, eberhard2004coarse, teodorovich2002renormalization, stepanyants2003effective, hristopulos2020random, colecchio2020multiscale}, reflecting the diversity of backgrounds such as volume averaging, homogenization, multiple-scale expansion methods, and stochastic perturbation theory.\\

In order to model flow in porous network models or fractured media, \replaced{which}{that} correspond to extremely contrasted local flow properties, the framework of percolation theory is more \replaced{suitable}{adapted} \citep{king1989use,berkowitz1993percolation,sahimi2011flow,sahimi1994applications,renard2013connectivity,hunt2014percolation,maillot2016connectivity,colecchio2021equivalent}. Karstic aquifers are even more extreme and complex underground multiscale structures\added{,} that include caves, sinkholes, and fracture networks. Modeling groundwater motion within such systems is a major societal issue in the context of overall climate change. The increasing occurrence of extreme events (droughts, floods) justifies building predictive models that remain robust even when forced by extreme data \citep{HALIHAN1998,Naughton2017,Mayaud2019}. This requires constructing a model that captures the main subsurface processes involved in such systems while representing their internal geometrical complexity as accurately as possible. It is appealing to adapt a model based on a network of conduits that can be characterized through direct exploration, staying close to the so-called distributed model approach. It differs from reservoir models and neural networks \citep{Jeannin2013,JEANNIN2021126508}\added{,} the aim of which is to relate input data (recharge) to rates at springs without referring to an explicit knowledge of the subsurface. The model is calibrated through data history, \replaced{working}{so working} well in generic recharge conditions, but accuracy may be lost if extreme forcing conditions are not present in the calibration data basis.

A natural representation of these structures can be achieved using a graph-theoretical framework as proposed by~\cite{GOUY2024130878} and references therein, which captures the interconnections of different elements. To describe fluid flow in these networks, assuming they are saturated and within the linear (Darcean) assumptions, they can be modeled as a Random Resistor Network (RRN). These RRNs can be mathematically described using a graph or a lattice\added{,} where each connection, or edge, between vertices $i$ and $j$ (points in the lattice) is assigned an element that can be assimilated to a resistor with a conductivity $T_{ij}$ that does not depend on the local potential difference, up to a first approximation.

In this study, we explore the interplay between the geometry of the underlying network and the conductivity distribution, and how both types of randomness influence the effective conductivity of RRNs. Before using more formal definitions, the effective conductivity can be defined as the scaling factor that matches the response of the heterogeneous network to that of its ``average'' counterpart under a given large-scale forcing. The imposed forcing may correspond to the usual forced flow in a given direction, as provided by homogenization and flux-averaging theories. It can also correspond to imposed boundary conditions such as recharge areas, catchments, etc. This allows the definition of a so-called effective conductivity that summarizes the overall behavior of the RRN and may help in devising a reduced-order model described by \replaced{fewer}{less} degrees of freedom. The situation may be more subtle when starting from a purely network approach that does not refer to any underlying Euclidean metric, precluding the idea of an imposed large-scale head gradient. The corresponding fluxes and potential drops are studied to highlight localization phenomena (channeling)\added{,} where a small subset of the network controls the overall flow properties, carrying most of the flow.

As stated above, two kinds of disorder may coexist: geometrical disorder, encapsulated in the so-called adjacency matrix of the graph that summarizes the connections between vertices, and conductivity disorder, which describes the various conduit \replaced{conductivities}{conductivity} that may be encountered. In the porous network or karst case, these values depend on network sub-scale details\added{, such as} the precise pore or conduit geometry \citep{frantz2021}. As a first approximation, these can be considered positive random variables, assumed to be independent \replaced{of}{on} the network geometry.
Considering a uniform conductivity distribution with a value $T_0$, the dominant descriptor is an order-parameter $p$ that controls the overall connectivity of the network. It is typically the proportion of active bonds (or edges in a graph-theoretic language). The framework offered by percolation theory is well-known among the physics and hydrology communities. The critical behavior and localization phenomena leading to critical path analysis can be found in review papers and textbooks \citep{berkowitz1993percolation,sahimi2011flow,Adler20131,hunt2014percolation}. Investigations in this area have been carried out by researchers interested in flow through Discrete Fracture Networks (DFN)\added{,} such as \citep{deDreuzy2012,noetinger2012quasi,Hyman2019}. This field has long incorporated graph theory to model flow in DFNs.\\

On the other hand, if the geometric disorder is low and the randomness of the conductivity values dominates, it is common practice, particularly within the geoscience community, to use upscaled numerical models. The discretized flow equations are solved using a simulator that uses a coarse grid\added{,} which smooths the fine-scale details of the input conductivity map. In such approaches, it is customary to determine coarse grid properties from their corresponding high-resolution counterparts provided by geologists \citep{noetinger2023random} using simplified approaches. Many techniques exist, as presented in review papers on the subject already cited above. Among these techniques\added{,} we focus on power averaging \citep{matheron1967,journel1986power,desbarats1992spatial,neuman1993prediction}. It allows quick characterization of the effective conductivity of heterogeneous formations by spatially averaging the conductivity field over a volume that may correspond to any coarse grid block employed by flow simulators\added{.} Its expression is given by:
\begin{eqnarray}
K^{eff} &=& \langle K^{\omega}\rangle^{\frac{1}{\omega}}\label{Eq:powerave}
\end{eqnarray}
Here, $\omega$ is an exponent such that $-1 \leq \omega \leq 1$, which depends on the dimension $D = 1, 2, 3$ of the space and on the spatial arrangement of the conductivity. Upon computing the limit, the value $\omega = 0$ corresponds to the geometric mean $K_g = \exp(\langle \log(K) \rangle)$. This provides a practical way to approximate conductivity upscaling, even though its theoretical foundations are mainly limited to specific models and conditions. Numerical tests show that power averaging is quite robust, even for \replaced{large values}{large value} of the log-conductivity variance \cite{neuman1993prediction,colecchio2020multiscale,colecchio2021equivalent}.\\

Fewer studies exist where both disorders are present \citep{Charlaix1987,deDreuzy2010}. Such combined models are more realistic since most natural conductive systems may be well-connected but not necessarily close to any percolation threshold, \replaced{implying they must be studied on a more case-by-case basis}{implying to be studied using more case by case basis} because the so-called universal results of percolation theory may be lost. In many cases, local conductivity variability may compete with \replaced{the effect of connectivity}{connectivity effect} on large-scale flow behavior. In situations where it dominates, many studies and classical textbooks address this topic in the hydrology or oil and gas literature \citep{matheron1967,renard1997calculating,sanchez2006representative,dagan2013,noetinger2013explicit, colecchio2020multiscale}. The basic concepts involve averaging and upscaling models that provide solutions to the following issue: how to propagate heterogeneity from the fine scale to the overall scale of the system, which is more relevant for applications. In these fields, simulations are carried out using flow simulators that solve discrete forms of the conservation equations. Great care is taken to account for boundary conditions at aquifer boundaries or at wells, if present. In this context, current practice in upscaling mainly involves grid coarsening and pragmatic solutions to the following issue: how to coarsen the mesh while accounting for as many sub-coarse grid details as possible, which can drastically change the overall flow behavior.\\

\replaced{In the present paper, the}{In present paper that} interaction between geometry and heterogeneity is studied\added{,} considering model RRN's described by a proportion of active edges controlling the overall connectivity, and a heterogeneitydescribed by means of a conductivity distribution. The former will be carried out considering model networks constructed from 2D or 3D lattices having various connectivity descriptors (typically a proportion $p$ of active edges), using conductivity distributions following a log-normal law.\\

The paper is organized as follows. In \replaced{the}{next} Sec.~\ref{subsec:The model}, the model and notations are introduced. Then in Sec.~\ref{subsec:averaging}, we develop a perturbation theory with respect to the conductivity fluctuation variance, starting from an arbitrary average connected network. \replaced{The overall framework of percolation theory is not presented;}{Percolation theory overall framework is not presented,} we refer to classical textbooks. \replaced{This}{That} allows us to set up some connections between percolation and perturbation theory. The basic approach is to average over the conductivity disorder at fixed network geometry. The aim is to answer the following question: is it possible to replace RRN's with random conductivities by another so-called equivalent RRN characterized by a so-called equivalent conductivity that may be computed from the local conductivity distribution using e.g.\ a power averaging formula? Such an extrapolation of second-order perturbation results may be justified using a mean-field argument presented in Appendix~\ref{App:meanfield}. Another approach\added{,} in which an overall effective conductivity is determined such that the response of the full RRN to an overall forcing \replaced{is}{are} similar\added{,} this closer to standard definitions of effective parameters used by hydrologists. As will be shown in Sec.~\ref{sec:secondorder}, it is possible to achieve this up to the second order while retaining the same underlying network of active edges. A similar mean-field argument presented in Appendix~\ref{App3} allows us to present the result under a more robust power averaging form valid for higher variances. \replaced{This}{That} allows us to discard the conductivity disorder, focusing on overall connectivity issues. Upscaling methods \replaced{that seek a sparsified network with fewer degrees of freedom}{where it is looked for a sparsified network involving less degrees of freedom} are not addressed in the present paper and will be the topic of another study.

In the next Sec.~\ref{sec:numerical-methodology}, the numerical methodology is presented, including RRN generation procedures that were carried out. Then\added{,} numerical results are treated within the proposed theoretical framework in Secs.~\ref{sec:resultsDdimensionalsructuredRRN} and \ref{sec:results}. It is shown that replacing the RRN with random conductivities by an ``homogeneous'' one with rescaled conductivities by means of a power law formula can be efficient, even for quite large log-conductivity variance\added{s} for which perturbation theory results \replaced{break down}{breakdown}.\\

\section{\label{sec:Theoryl}Theoretical considerations}
\subsection{\label{subsec:The model} Model and notations}
We consider a RRN represented by a connected undirected graph of $N$ vertices $G = (V, E)$. To each edge $<i,j> \in E$ connecting vertices $i$ and $j$, a conductivity $T_{i j} = T_{j i} \neq 0$ is selected randomly \replaced{from}{on} a distribution of positive real numbers of pdf $p_{i j}$. They are assumed to be independent \replaced{of}{on} each other. The number of edges is $N_E$.\\ 

We consider the set of potentials $\mathbf{P}=(P_1,...,P_N)^\top$ \replaced{that solve}{solution of} the linear system corresponding to solving Kirchhoff's laws:
\begin{eqnarray}
\forall i \sum _{j \in <i>} T_{i j}(P_j-P_i)= Q_i \label{eq:basicA}\\
\sum _{i} P_i=0
\label{eq:basic}
\end{eqnarray} 
The last equation allows \replaced{us to obtain}{to get} a well-defined set of equations, as potentials are defined up to an arbitrary additive constant.

Similarly, the source terms $Q_i$ are such that $\sum_{i} Q_{j} =0$. The set of labels $<i>$ denotes the set of vertices connected to vertex $i$ by one edge. As the graph is connected, the normalization conditions for both $\mathbf{P}$ and $\mathbf{Q}$ \replaced{ensure}{give} a well-defined unique solution. For short, this linear system may be written $\mathbf{L \cdot P=Q}$, in which $\mathbf{L}$ summarizes the corresponding matrix of the linear system Eqs.~\ref{eq:basicA} and ~\ref{eq:basic}. \\

\subsection{\label{subsec:averaging}Averaging procedure}
In most practical cases, we are interested in the average solution $ \bar P_i = \langle P_i \rangle$, which unfortunately is not the solution of the average set of equations. The average $\langle \dots \rangle$ is to be evaluated over all the set of $T_{i j}$'s weighted by their respective pdf's. More specifically, we look for an effective equation relating $ \bar P_i $ to $\mathbf{Q}$. So, we seek whether $ \bar P_i $ satisfies an effective set of averaged equations of the form:
\begin{eqnarray}
\forall i,\quad \sum _{j} T^{eff}_{i j}(\bar P_j-\bar P_i)= Q_i \\
\sum _{i} \bar P_i = 0  \label{Eq:sumcondition}
\label{eq:defaverage}
\end{eqnarray}

Note that in \replaced{this}{that} equation, it is not assumed that summations over $j$ are restricted to the initial set of vertices $<i>$ incident to $i$ in the starting point problem. In addition, it is implicitly assumed that the average potentials follow a Kirchhoff-like set of equations\added{,} \replaced{which}{that} will be justified \replaced{shortly}{right now}. In formal terms, we are investigating the following matrix $\mathbf{L}^{eff}=\langle \mathbf{L}^{-1} \rangle ^{-1}$. Although at first sight $\mathbf{L}$ is not invertible, the condition \ref{Eq:sumcondition} combined with the no net source term condition $\sum_{i} Q_{i} = 0$ \replaced{gives meaning to}{gives sense to} the inverse in the relevant subspace. It can be observed that $\mathbf{L}^{eff}$ is obviously symmetric, being the \replaced{inverse}{invert} of a symmetric matrix.\\

In order to check that $\mathbf{L}^{eff}$ possesses the structure of a Laplacian matrix, it is useful to decompose $\mathbf{L} $ under the form $\mathbf{L} =\mathbf{L_0} + \delta \mathbf{L}$. The matrix $\mathbf{L_0}$ is determined using average conductivities $\langle T_{ij} \rangle$, and $\delta \mathbf{L}$ corresponds \replaced{to}{thus to} the fluctuating part of the Laplace operator written using $\delta T_{ij}$ of \replaced{zero}{null} average. Thus, $\mathbf{L_0}$ corresponds to the average set of equations.\\

Writing $(\mathbf{L_0} + \delta \mathbf{L})^{-1} = \mathbf{L_0}^{-1} (\mathbf{1} + \delta \mathbf{L}\mathbf{L_0}^{-1})^{-1}$, \replaced{we obtain}{one obtain}: $\mathbf{L}^{eff}=\langle  (\mathbf{1} + \delta \mathbf{L}\mathbf{L_0}^{-1})^{-1} \rangle ^{-1} \mathbf{L_0}$. Assuming that $\mathbf{M}=\langle  (\mathbf{1} + \delta \mathbf{L}\mathbf{L_0}^{-1})^{-1} \rangle ^{-1}$ is a well-defined matrix, $\mathbf{L_0}$ is a Laplacian matrix such that $\sum_{j} L_{ij}=0$, so a similar property may be derived for $\mathbf{L}^{eff} = \mathbf{M} \cdot \mathbf{L_0}$ because $\mathbf{L}^{eff}_{ij} = \sum_{k} M_{ik} L_{kj}$, so $\sum_{j}\mathbf{L}^{eff}_{ij} = \sum_{k}\sum_{j} M_{ik} L_{kj} = 0$. As $\mathbf{L}^{eff}$ is symmetric, gathering these equalities \added{confirms} that $\mathbf{L}^{eff}$ has the structure of a Laplacian matrix leading to an effective set of equations sharing the form \ref{eq:defaverage}. \replaced{Note}{Notice} that this result is quite general and does not rely on a specific conductivity distribution nor on any order of a perturbation expansion.\\

Writing $(\mathbf{L_0} + \delta \mathbf{L})^{-1} = \mathbf{L_0}^{-1} (\mathbf{1} + \delta \mathbf{L}\mathbf{L_0}^{-1})^{-1}$, \replaced{we obtain}{one obtain}: $\mathbf{L}^{eff}=\langle  (\mathbf{1} + \delta \mathbf{L}\mathbf{L_0}^{-1})^{-1} \rangle ^{-1} \mathbf{L_0}$. Assuming that $\mathbf{M}=\langle  (\mathbf{1} + \delta \mathbf{L}\mathbf{L_0}^{-1})^{-1} \rangle ^{-1}$ is a well-defined matrix, $\mathbf{L_0}$ is a Laplacian matrix such that $\sum_{j} L_{ij}=0$, so a similar property may be derived for $\mathbf{L}^{eff} = \mathbf{M} \cdot \mathbf{L_0}$ because $\mathbf{L}^{eff}_{ij} = \sum_{k} M_{ik} L_{kj}$, so $\sum_{j}\mathbf{L}^{eff}_{ij} = \sum_{k}\sum_{j} M_{ik} L_{kj} = 0$. As $\mathbf{L}^{eff}$ is symmetric, gathering these equalities \added{confirms} that $\mathbf{L}^{eff}$ has the structure of a Laplacian matrix leading to an effective set of equations sharing the form \ref{eq:defaverage}. \replaced{Note}{Notice} that this result is quite general and does not rely on a specific conductivity distribution nor on any order of a perturbation expansion.\\

That procedure corresponds physically to most averaging techniques employed in \replaced{this}{that} class of issues, in which the source terms\added{,} such as boundary conditions\added{ and} imposed potentials\added{,} are known \replaced{at a few points}{on few points}, and internal degrees of freedom of the bulk that follow conservation equations are to be eliminated from the problem, resulting in solving local equations. In analogy with continuous Laplace equation averaging \cite{noetinger1998use}, it can be anticipated that $\mathbf{L}$\added{,} being a sparse matrix having ($N \le N_{non\ zero \ entries} \ll N^2$ \replaced{non-zero}{non zero} elements), $\langle \mathbf{L}^{-1} \rangle$ can be expected to be a full matrix of $N^2$ elements (reflecting the long-ranged character of the Laplace equation Green's function). But it can also be anticipated that $\langle \mathbf{L}^{-1} \rangle^{-1}$ may be sparse, or have exponentially small matrix elements far from the diagonal (although in the context of graphs, a more precise meaning should be given to \replaced{this}{that} statement in terms \replaced{of}{on} an intrinsic distance on the graph)\added{,} such as the chemical distance between nodes. \replaced{This}{That} may be justified \added{by} studying carefully the matrix $\mathbf{M}=\langle  (\mathbf{1} + \delta \mathbf{L}\mathbf{L_0}^{-1})^{-1} \rangle ^{-1}$. 

Extrapolating results obtained in the continuous case context \cite{noetinger1998use}, it can be observed that $\mathbf{M}$ may be similar to the inverse Fourier transform of $\dfrac{1}{1+(qa)^2+\cdots}$, in which $\cdots$ represent higher powers of $q$. Coming back to the real domain, \replaced{this}{that} provides an operator having exponentially decreasing elements in the real domain. Having a direct and rigorous derivation in the graph-theoretic context would be useful.\\

In the case of the continuous Laplace equation, it can be shown \citep{noetinger1998use} that for an infinite system, such a definition coincides with the usual, more intuitive definitions involving estimations of the mean flux as a response to an imposed head gradient. Following \replaced{this}{that} procedure, effective conductivities $T_{ij}^{eff}$ are to be estimated from the input data, which requires the use of perturbation expansion methods to obtain explicit results.

\subsection{\label{sec:secondorder} Second order perturbation expansion results}
In order to set up a perturbation expansion, we write $(\mathbf{L_0} + \delta \mathbf{L})^{-1} = \mathbf{L_0}^{-1} (\mathbf{1} + \delta \mathbf{L}\mathbf{L_0}^{-1})^{-1}$\added{,} then the second factor may be expanded in a Neumann series\added{,} \replaced{which is equivalent to solving}{that is equivalent to solve} the equations by an iterative geometric sequence. After some manipulations\added{,} we obtain:
\begin{eqnarray}
\langle \mathbf{L}^{-1} \rangle ^{-1}=[\mathbf{L_0}^{-1}\langle (\mathbf{1}+\delta \mathbf{L}\mathbf{L_0}^{-1})^{-1}\rangle]^{-1}\nonumber \\
=\mathbf{L_0}[\mathbf{L_0}+\sum _{n=1}^{\infty}(-1)^{n}\langle(\delta \mathbf{L}\mathbf{L_0}^{-1})^{n}\rangle\mathbf{L_0}]^{-1}\mathbf{L_0}
\label{eq:Neuman}
\end{eqnarray}

The averaging process can be carried out using \replaced{this}{that} series expansion, but due to the inversion of the series of matrices between brackets in Eq.~\ref{eq:Neuman}, \replaced{the computation of high-order terms quickly becomes cumbersome}{computation of high order terms becomes quickly cumbersome}. In the continuous case, considering a D-dimensional random conductivity problem, some general results \replaced{allowing us to obtain}{allowing to get} an alternative form of a series expansion of $\langle \mathbf{L}^{-1} \rangle^{-1}$ in terms of a series involving so-called 1P irreducible Feynman diagrams containing correlation functions of conductivity fluctuations of increasing order can be \replaced{derived}{obtained} \citep{noetinger1998use}. \replaced{This}{That} summation \replaced{bypasses}{by-passes} the inversion of the sum of a series of matrices between brackets in \replaced{Equation~\ref{eq:Neuman}}{equation \ref{eq:Neuman}}. \replaced{This confirms}{That confirms} the fact that $\langle \mathbf{L}^{-1} \rangle^{-1}$ shares essentially the same structure \replaced{as}{than} the original average Laplace operator $\mathbf{L_0}$.\\

We were not able to derive such a summation formula in the present discrete case, so \replaced{for now}{right now} the analysis is limited to the second-order expansion\added{,} \replaced{which is}{that is} given by:

\begin{eqnarray}
\langle \mathbf{L}^{-1} \rangle ^{-1}=[\mathbf{L_0}^{-1}\langle (\mathbf{1}+\delta \mathbf{L}\mathbf{L_0}^{-1})^{-1}\rangle]^{-1}\nonumber \\
=\mathbf{L_0}[\mathbf{L_0}+\sum _{n=1}^{\infty}(-1)^{n}\langle(\delta \mathbf{L}\mathbf{L_0}^{-1})^{n}\rangle\mathbf{L_0}]^{-1}\mathbf{L_0}
\label{eq:Neuman2}
\end{eqnarray}

The matrix $\mathbf{\Sigma} \simeq \langle \delta \mathbf{L}\mathbf{L_0}^{-1} \delta \mathbf{L} \rangle$ is called the self-energy in condensed matter physics \citep{noetinger1998use}.

The net second\replaced{-order}{ order} result can be evaluated\added{,} by computing the product $\mathbf{\Sigma} \cdot \mathbf{P}$:

Using the equality $\langle \delta T_{ik} \delta T_{lj} \rangle = \sigma_{ik}^2 [\delta_{il} \delta_{kj} + \delta_{ij} \delta_{kl}]$\added{,} one gets:

\begin{eqnarray}
(\mathbf{\Sigma} \cdot \mathbf{P})_{i} = \sum_{j \in <i>} \sigma_{ij}^2 [{L_0}^{-1}_{ij}+{L_0}^{-1}_{ji}-{L_0}^{-1}_{ii}-{L_0}^{-1}_{jj}](P_{j}-P{_i})
\label{Eq:secondorder}
\end{eqnarray}

It appears that \replaced{up to}{up at} second order, the matrix $\langle \mathbf{L}^{-1} \rangle ^{-1}$ couples the same vertices \replaced{as}{than} the average input $\mathbf{{L}_0}$, so the associated connectivity matrix is equal to the initial one up to that order. The symmetry of $\mathbf{\Sigma}$ appears clearly\added{,} as it should. The quantity $R_{ij} = [{L_0}^{-1}_{ij} + {L_0}^{-1}_{ji} - {L_0}^{-1}_{ii} - {L_0}^{-1}_{jj}]$ has a sound physical interpretation: it corresponds to the potential difference between vertices \added{$i$ and $j$} computed using the average RRN\added{,} given that a unit-strength source/sink dipole is located on vertices \added{$i$ and $j$}. In graph theory\added{,} it corresponds to the so-called resistance distance introduced by \cite{Klein1993ResistanceD}, a major topic in \replaced{the}{modern} graph theory context \cite{spielman2008graph}. The edge effective conductivity $T_{ij}^{eff}$ can be estimated up to second\replaced{-order}{-order} from that formula. In order to \replaced{obtain}{get} a more robust estimator capturing higher variance, a mean-field theory is presented in Appendix~\ref{App:meanfield}\added{,} \replaced{which justifies}{that allows to justifies} the power-average extrapolation Eq.~\ref{Eq:powerave}\added{,} sharing the same second-order expansion.

\subsection{\label{sec:resultsDdimensionalsructuredRRN} Results for structured RRN, percolation networks}
The preceding calculations can be specialized for $D$-dimensional structured networks\added{,} such as those resulting from the discretization of a Laplace equation in a $D$-dimensional space using finite differences with a $2D+1$ stencil. In such a case, the corresponding RRN equations read: 
$$\sum _{\epsilon_{i{_k=}} \pm 1}T_{i_{1}, \dots i_{D};i_{1}+\epsilon _{i_{1}}, \dots i_{D}+\epsilon _{i_{D}}}(P_{i_{1}+\epsilon _{i_{1}}, \dots i_{D}+\epsilon _{i_{D}}}-P_{i_{1}, \dots i_{D}})= Q_{i_{1}, \dots i_{D}}$$

The average matrix $\mathbf{L_0}$ is symmetric\added{;} it has $2D+1$ diagonals, with the form 
$$ \langle T \rangle (0\cdots 1, 1,.., -2D, 1,1,1,0 \cdots 0)$$ 
with $2D$ \replaced{off-diagonal unit entries}{$\times$ unit values off diagonal}, and $-2D$ on the diagonal.

Assuming a uniform conductivity variance $\sigma^2$\added{,} one obtains in the limit of a large network, Appendix~\ref{App3}:
\begin{eqnarray}
(\mathbf{\Sigma} \cdot \mathbf{P})_{\bf i}= \sum_{{\bf j} \in <{\bf i}>} \frac{\sigma^2}{\langle  T \rangle D} (P_{{\bf j}}-P{_{\bf i}})
\label{Eq:secondorderDdim}
\end{eqnarray}

Here ${\bf i}$ (resp. ${\bf j}$) denotes the multi-index ${\bf i}=(i_{1}, \dots, i_{D})$ (resp. ${\bf j}=(j_{1}, \dots, j_{D})$). For $D=1$, considering a path graph such as those considered in Appendix~\ref{apppathgraphs}, the formula is exact without any large-$N$ assumption. In other cases, assuming a large RRN\added{,} and that the statistical properties of $T_{ij}$ do not depend on the edge $ij$\added{,} it can be shown that up to second order, the system behaves as an effective RRN with conductivities:

\begin{eqnarray}
T^{eff}_{\bf ij} &=& \langle T_{\bf ij} \rangle \left(1-\frac{1}{D}\frac{\sigma^2}{\langle  T_{\bf ij} \rangle^{2}}+\cdots\right)\nonumber \\
&\simeq& \langle T_{g \bf ij}\rangle \exp (1/2-1/D)\sigma_{\log T}^2 = \langle T_{\bf ij}^{(1-2/D)}\rangle^{\frac{1}{(1-2/D)}}.\label{Eq:LLMD}
\end{eqnarray}

Here, $T_{g \bf ij} = \exp \langle \log T_{\bf ij} \rangle$ and $\sigma_{\log T}^2$ are respectively the geometric average of the edge $ij$ conductivity and the associated log-conductivity variance. A mean-field argument \replaced{that justifies}{allowing to justify} \replaced{this}{that} extrapolation is given in \ref{App:meanfield}. We used textbook formulas for estimating power averages of log-normal distributions. \replaced{This last form, which shares the same second-order expansion, is known as the Landau-Lifshitz-Matheron conjecture}{This last form that shares the same second order expansion is a result known under the name of Landau-Lifshitz-Matheron conjecture} \citep{landau1960lifshitz,matheron1967,n1994effective}, \replaced{which was}{that was} derived in the context of continuous Laplace problems in heterogeneous media. \replaced{The}{That} conjecture is exact at all orders for $D=1$\added{,} in which a complete analytical solution is available (Appendix~\ref{apppathgraphs}). In 2D, it is exact for specific log-normal distributions or \replaced{near}{close to} percolation thresholds \citep{matheron1967, stauffer2014introduction}. In the 3D continuous case, it was shown to be inexact using sixth-order expansion \citep{de1995correlation,abramovich1995effective,stepanyants2003effective}\added{,}  although it was observed to be very accurate in numerical practice \citep{neuman1993prediction,romeu1995calculation,wang2017finite}.
\\

In cases where connections are kept with probability $p$ close to 1 (full network), an estimation of $\omega$ is given by a quite simple formula (\added{with} derivation in Appendix~\ref{App3}).

 \begin{equation}
\omega \approx 1-2*\frac{N_{\text{p}}}{N_{\text{E}}}
 \label{eq:NsNb}
\end{equation}
Here, $N_{\text{E}}$ represents the number of active edges\added{:} the set of edges that are available to flow. $N_{\text{p}}$ denotes the number of vertices connected to these edges (active vertices). The equation can be expressed in terms of $k$, the average degree of a node in the graph, where $k = \frac{2N_{\text{E}}}{N_{\text{p}}}$\added{;} thus\added{,} $\omega = 1 - \frac{4}{k}$. 

This expression satisfies two conditions: 
(1) As $p \to 1$\added{,} it yields the value of $\omega$ known for 2D and 3D full networks with a lognormal conductivity distribution, where $\omega = 0$ for 2D and $\omega = \frac{1}{3}$ for 3D networks; and 
(2)\deleted{,} when $\frac{N_{\text{p}}}{N_{\text{E}}} = 1$\added{,} it results in $\omega = -1$, the exact value for 1D networks (corresponding to the harmonic mean). A 1D backbone network may behave as a 1D network when $p = p_c$.

\section{Numerical Methodology}
\label{sec:numerical-methodology}
This section details the methodology used to generate Random Resistor Networks (RRNs) for investigating the impact of conductivity and geometric disorders on the effective conductivity, $T^{eff}$. In these networks, geometric disorder is introduced through the random removal of edges, while conductivity disorder is implemented by varying the conductivity values assigned to the edges.

\subsection{\label{sebsec:GenRRN}Generation of RRNs}
\subsubsection{\label{subsubsec:percolationLattice}Generation of 2D and 3D lattice RRN with r=1}

\begin{figure}[h!]
    \centering

   \includegraphics[width=0.48\linewidth]{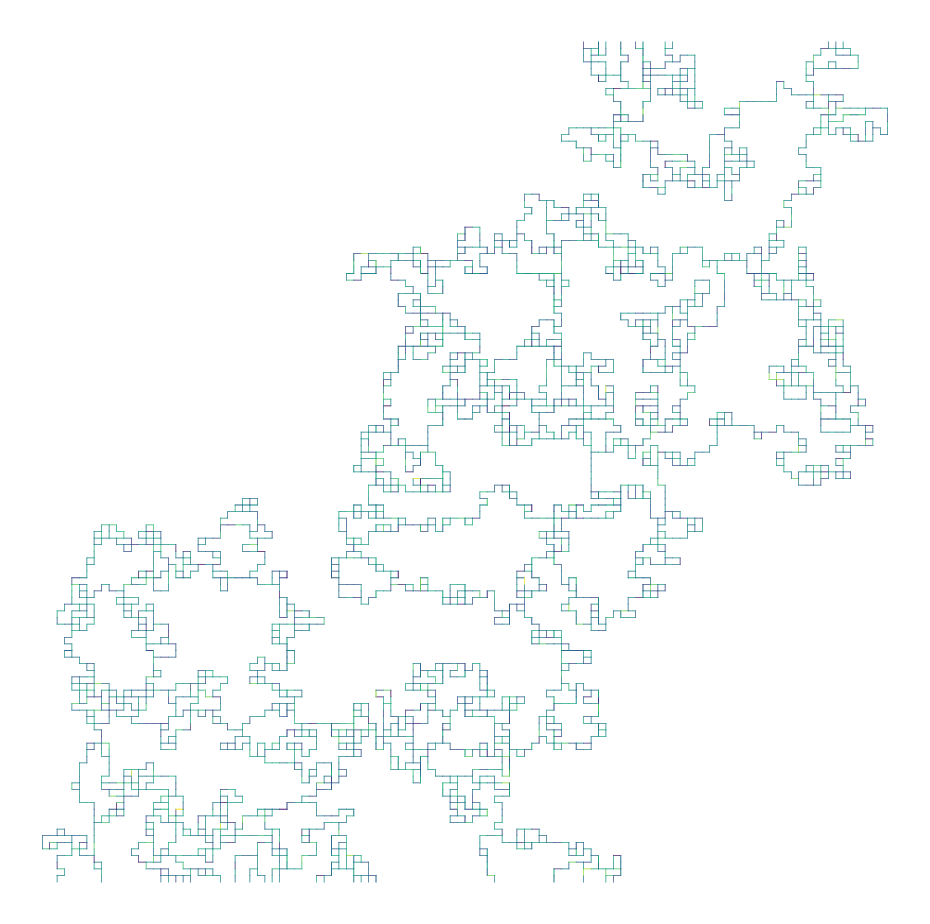}  
   \includegraphics[width=0.48\linewidth]{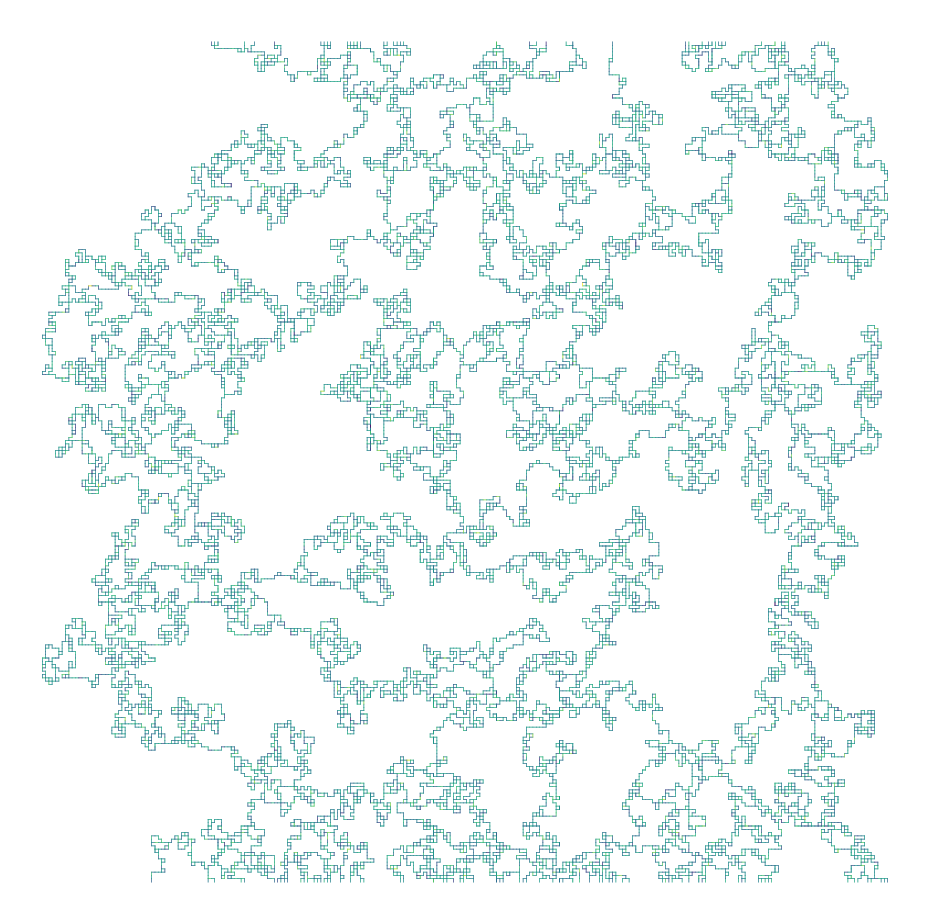}

    \caption{Networks at the critical degree ($p \approx p_c$), with 2D networks ($128 \times 128$ nodes) on the left and 3D networks ($48 \times 48 \times 48$ nodes) on the right. Edge colors represent conductivity values.}
    \label{fig:plot_net_perco}
\end{figure}

Nodes are placed on a regular square (2D) or cubic (3D) grid. In this configuration, each node is connected only to its immediate orthogonal neighbors (excluding diagonal connections)\added{,} to emulate a lattice structure commonly explored in classical percolation theory \citep{stauffer2014introduction}. This connectivity setup, defined by a radius $r = 1$, ensures that nodes interact only with their nearest orthogonal neighbors. \replaced{The geometry is varied}{Geometry is varied} by selectively removing edges, effectively setting their conductivity to zero. The connectivity proportion $p$ directly dictates the probability of an edge being retained, simulating different scenarios from sparse to fully connected networks as $p$ varies from the percolation threshold $p_c$ to 1. On the other hand, conductivities are assigned independently from a lognormal distribution with a unitary geometric mean, reflecting the natural variability of conduit sizes \citep{frantz2021}. The variance of the logarithm of this distribution, $\sigma^2_{\log T}$, controls the heterogeneity within the network. Examples of these network configurations can be seen in Fig.~\ref{fig:plot_net_perco}, which illustrates structured networks at the critical degree ($p=p_c$) near the percolation threshold.

\FloatBarrier

\subsubsection{\label{subsubsec:scalingPercolationRRN} Structured RRNs with $r>1$}

For networks where the connectivity radius $r$ extends beyond the immediate neighbors, we simulate a transition from local to global connectivity. The radius $r$ serves as a scaling parameter that, along with $p$, adjusts connectivity and shapes the network's structure. This scaling allows for the exploration of network behaviors across a continuum of connectivity regimes.

\begin{figure}[h!]
    \centering
        \includegraphics[width=0.48\linewidth]{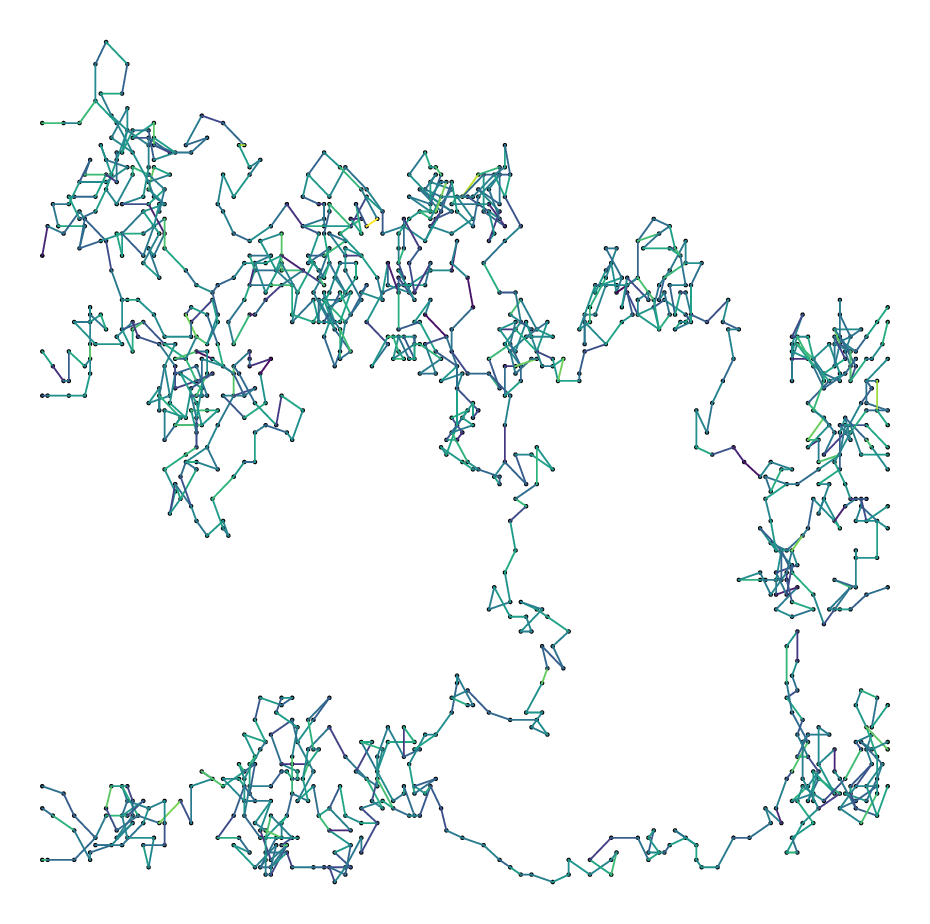}  
        \includegraphics[width=0.48\linewidth]{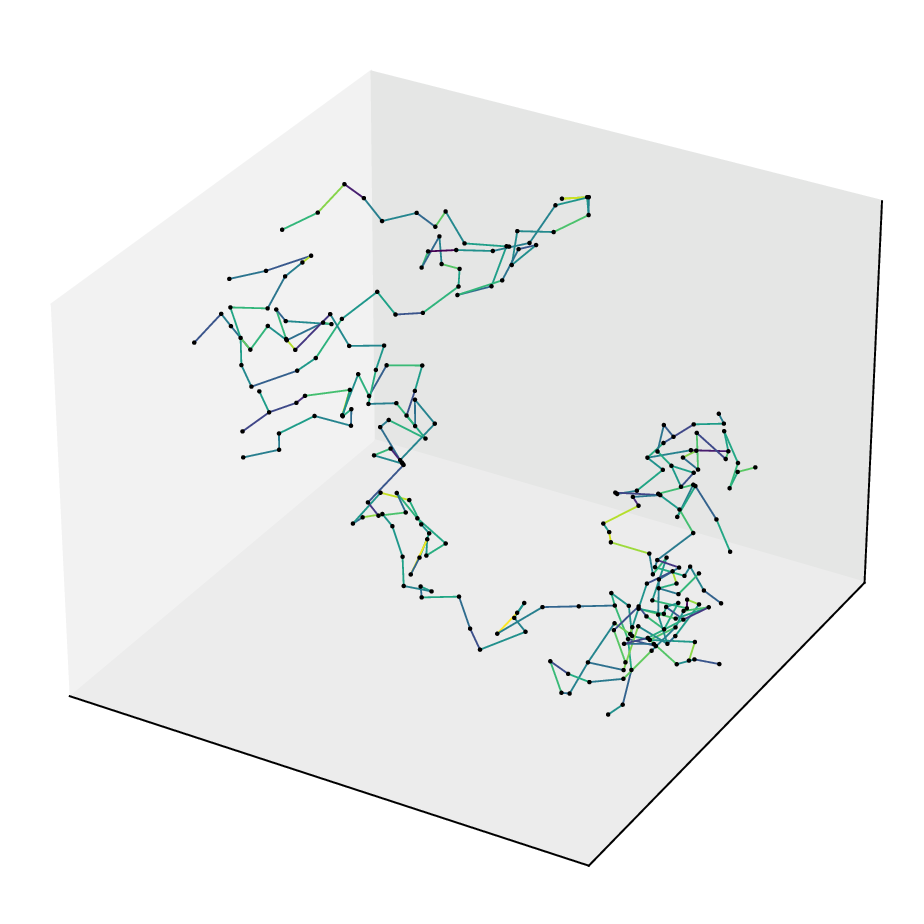}
    \caption{Examples of structured networks at critical connectivity ($p \approx p_c$): on the left, a 2D network with $200 \times 200$ nodes, and on the right, a 3D network with $48 \times 48 \times 48$ nodes. These networks incorporate a connectivity radius $r = 5$, that extends beyond nearest neighbors, altering the overall connectivity. Edge colors represent conductivity values.}
    \label{fig:plot_net_radius}
\end{figure}

The proportion $p$ is defined as $p = k / k_{r}$, where $k$ represents the average degree of connectivity after edge removal, and $k_{r}$ represents the average degree of a node in a network where all nodes within a distance $r$ from each other are connected. This degree depends on the radius $r$, as a larger radius increases the number of neighboring nodes that can be connected to each other. Similar strategies are employed to study ad hoc wireless networks under varying parameters of radius and connectivity \citep{Broutin2014}. Examples of these network configurations are visually represented in Fig.~\ref{fig:plot_net_radius}, which displays structured networks at connectivity near the critical degree ($k \approx k_c$) for a radius of $r=5$.


\FloatBarrier

\subsection{\label{sec:flowimul}Simulation of flow and postprocessing of the solution}

The potential at each node is computed by solving Kirchhoff's laws, as shown in Eq.~\ref{eq:basic}, with a potential difference $ \Delta P $ applied between the nodes in the inlet and the nodes in the outlet. In structured networks\added{,} where nodes are systematically arranged on a regular grid, the inlet is designated on one face of the domain\added{,} and the outlet is positioned on the opposite face.

The distance $ L $ between the inlet and outlet is defined as the number of edges $ N_x - 1 $. The cross-sectional area $ A $\added{,} perpendicular to the flow direction, is calculated as $ (N_y - 1) \times (N_z - 1) $. Here, $ N_x $, $ N_y $, and $ N_z $ represent the original number of nodes along each dimension before any removal of edges.

\replaced{The effective conductivity, $ T^{eff} $, encapsulates}{Effective conductivity, $ T^{eff} $, encapsulates} the hydraulic response of the network under a given large-scale forcing\added{,} and is computed by:

\begin{equation}
   T^{eff} = \frac{Q_T}{\Delta P} \frac{L}{A}
\end{equation}
where $ Q_T $ denotes the total flow.

To optimize the simulation process and minimize computational costs, a network pruning process \replaced{is applied before simulations begin}{is implemented prior to conducting flow simulations is applied before simulations begin} \citep{Knackstedt2000,Seiffe2022}. This process removes nodes of degree zero and one, which do not contribute to flow, and retains only nodes within the largest cluster. Following the pruning process, the linear system is solved to determine the potential at each node, leading to the computation of flow at each edge. This solution enables the computation of the number of active edges $N_{\text{E}}$ and active nodes $N_{\text{p}}$, and the calculation of the parameter $\omega$ using Eq.~\ref{eq:NsNb}.

In order to \replaced{gain}{get} a better understanding of the influence of both disorders \replaced{on}{about} the effective large-scale conductivity, the following procedure was employed. The main idea is to account for the conductivity disorder by modifying the power-law averaging method.

Firstly, for a lognormal distribution of conductivities, the power average of the input distribution is given by:

\begin{equation}
\langle T^{\omega}\rangle^{\frac{1}{\omega}}=T_g \exp{\left( \omega \frac{\sigma_{\log T}^2}{2}\right)}
\label{eq:powerAvLogn}
\end{equation}

Here, $T_g$ represents the geometric mean of the lognormal distribution, and $\sigma_{\log T}^2$ is the variance of its logarithm. For a specified set of parameters $\sigma_{\log T}^2, p$\added{,} an effective conductivity $T^{eff}(\sigma_{\log T}^2, p)$ can be calculated by solving the Kirchhoff equations. In the case of a fully connected RRN, the exponent $\omega$ is determined by matching this formula with the numerically obtained $T^{eff}$. This involves replacing the conductivity of each edge with a uniform value given by $T^{eff}$\added{,} resulting in an equivalent homogeneous RRN.

However, introducing topological disorder by removing edges alters the network's connectivity and the distribution of conductive pathways\added{,} which requires adjustments to the effective conductivity model to accurately reflect these changes. The modified effective conductivity can be determined using the following equation \citep{deDreuzy2010}:

\begin{equation}
T^{eff}( \sigma_{\log T}^2,p)=T^{eff}(0,p) \exp{\left( \omega \frac{\sigma_{\log T}^2}{2}\right)}
\label{eq:powerAvLognComplex}
\end{equation}

In this expression, $T^{eff}(0,p)$ represents the effective conductivity calculated for a network without conductivity disorder ($\sigma_{\log T}^2 = 0$)\added{,} but with geometric disorder. Therefore, flow simulations must be conducted twice: once without conductivity disorder ($\sigma_{\log T}^2 = 0$)\added{,} and once with conductivity disorder ($\sigma_{\log T}^2 > 0$)\added{,} to accurately determine the parameter $\omega$ for the network under study. The resulting parameter $\omega$\added{,} obtained by matching\added{,} will thus depend on $p$. The relevance of this procedure will be tested with the theoretical elements provided in Sec.~\ref{sec:secondorder}. Additionally, the validity of the power-law averaging must be evaluated\added{,} specifically its robustness against significant inputs of $\sigma_{\log T}^2$.

\FloatBarrier

\subsection{\label{subsec:simulpm}Setup of Simulation Parameters}
The structured networks are configured on grids with dimensions $N_x = N_y = 512$ for two-dimensional (2D) setups and $N_x = N_y = N_z = 64$ for three-dimensional (3D) setups. For the generation of network geometry and the assignment of conductivity values, variability is managed through seed-based Monte Carlo methods. Specifically, two separate seeds are used: one for generating the network geometry ($n_G$)\added{,} and another for the conductivity distribution ($n_C$).

\begin{figure}[h!]
    \centering
    \includegraphics[width=0.48\linewidth]{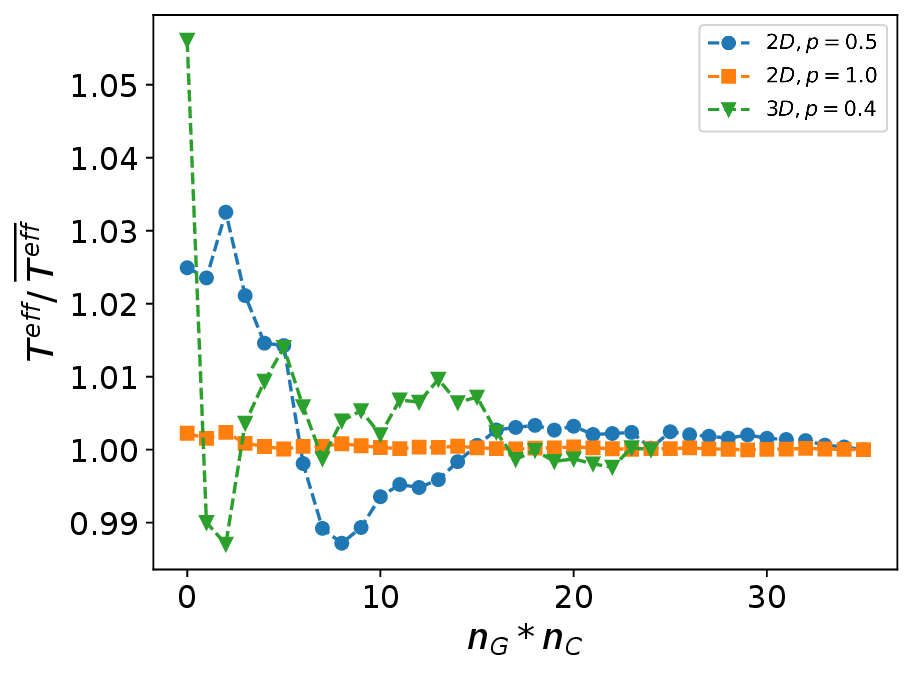}  
    \includegraphics[width=0.48\linewidth]{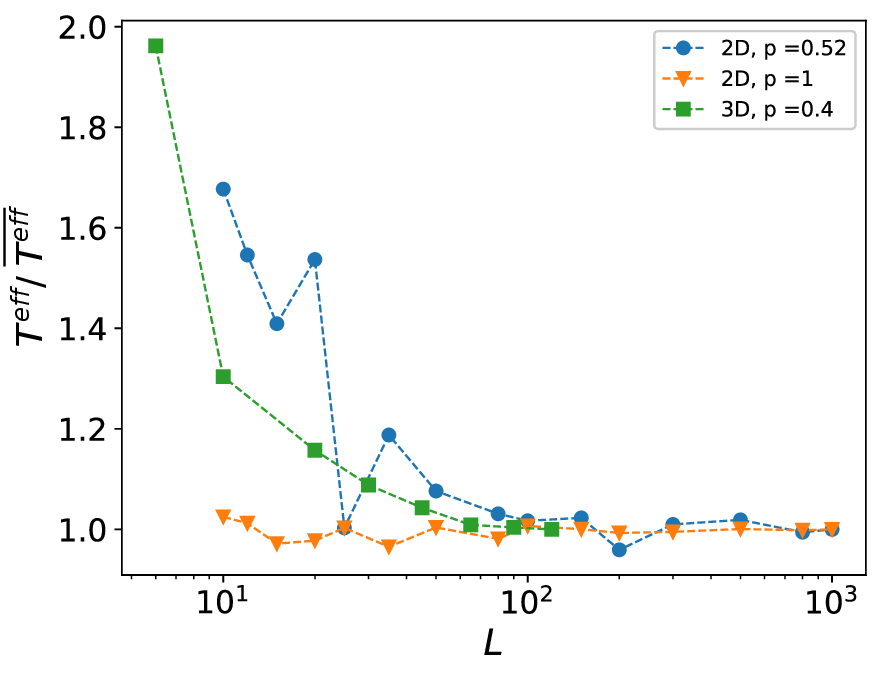}
    \caption{(Left) Influence of the number of realizations $N_{real} = n_C \times n_G$ on the average effective conductivity $T^{eff}$, normalized by $\overline{T^{eff}}$ for $N_{real} = 1000$, for structured networks of size $512 \times 512$ (2D) and $64 \times 64 \times 64$ (3D), and $\sigma^2_{\log T} = 5$. (Right) Dependence of $T^{eff}$ on network size $L$, normalized by $\overline{T^{eff}}$ for the largest $L$, with $r=1$ and $\sigma^2_{\log T} = 5$.}
    \label{fig:combined}
\end{figure}

The results, as shown in the left panel of Fig.~\ref{fig:combined}, indicate that a relatively small number of realizations are sufficient to achieve a relative error below 1\%. However, a large number of realizations were performed to ensure statistically representative results. A total of 900 realizations ($n_G \times n_C$)\added{,} with $n_G = 30$ and $n_C = 30$\added{,} were conducted to achieve statistical convergence of $T^{eff}$. This methodological choice enhances the robustness of the analysis, particularly near the percolation threshold, where fluctuations in $T^{eff}$ are expected \citep{colecchio2021equivalent,boschan2012scale}. Furthermore, for \replaced{values of the proportion of conductive edges $p$}{values of proportion of conductive edges $ p $} close to 1, the influence of geometric disorder becomes negligible, and statistical convergence is primarily governed by $n_C$.

Given that the effective conductivity $T^{eff}$ is notably dependent on the network size $L$, additional tests were conducted to identify the scale at which simulations should be executed to approximate the asymptotic value of $T^{eff}$. The findings are presented in the right panel of Fig.~\ref{fig:combined}. Based on these results, simulations were performed using network sizes of $L = 512$ for 2D configurations and $L = 64$ for 3D\added{,} to minimize size-dependent effects and ensure robust results.

To explore the effects of geometric disorder on $T^{eff}$, the proportion $p$ was adjusted from values close to the percolation threshold to those representing full connectivity. The variance of the lognormal distribution, $\sigma^2_{\log T}$, was varied from zero (where all conductivities are uniform) to a highly heterogeneous scenario with $\sigma^2_{\log T} = 5$. The simulation parameters are detailed in Table~\ref{table:parameters}.

\begin{table}[h!]
\centering
\caption{Simulation Parameters}
\label{table:parameters}
\begin{tabular}{|c|c|c|}
\hline
\textbf{Parameter} & \textbf{Description} & \textbf{Values} \\
\hline
$N_x, N_y, (N_z)$ & Dimensions  & 512 (2D), 64 (3D) \\
\hline
$r$ & Radius & [1,..,5] \\
\hline
$p$ & Proportion of conductive edges & [$p_c$,..,1] \\
\hline
$\sigma^2_{\log T}$ & Variance of log conductivity & [0,.., 5] \\
\hline 
$N_{real}$ & Total number of realizations  & 900 \\
\hline
\end{tabular}
\end{table}

\FloatBarrier

\section{Results\label{sec:results}}

\subsection{\label{subsec:2D2Dresultsr=1l}Effective conductivity and power average exponent of 2D and 3D lattice RRN with r=1}

\begin{figure}[h!]
  \centering
  \includegraphics[width=0.48\textwidth]{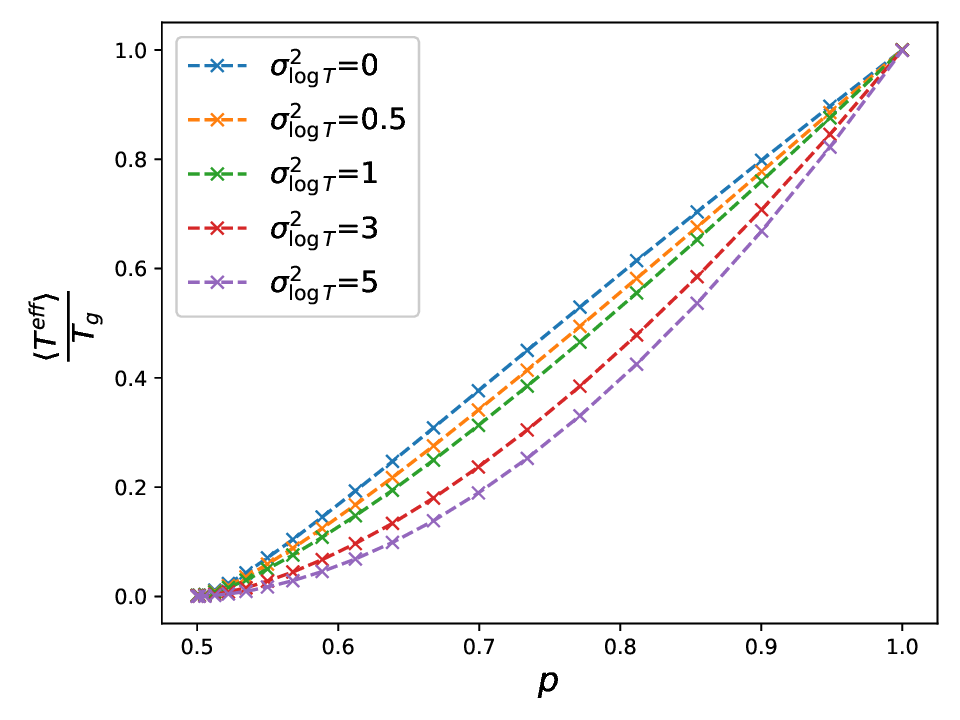}
  \includegraphics[width=0.48\textwidth]{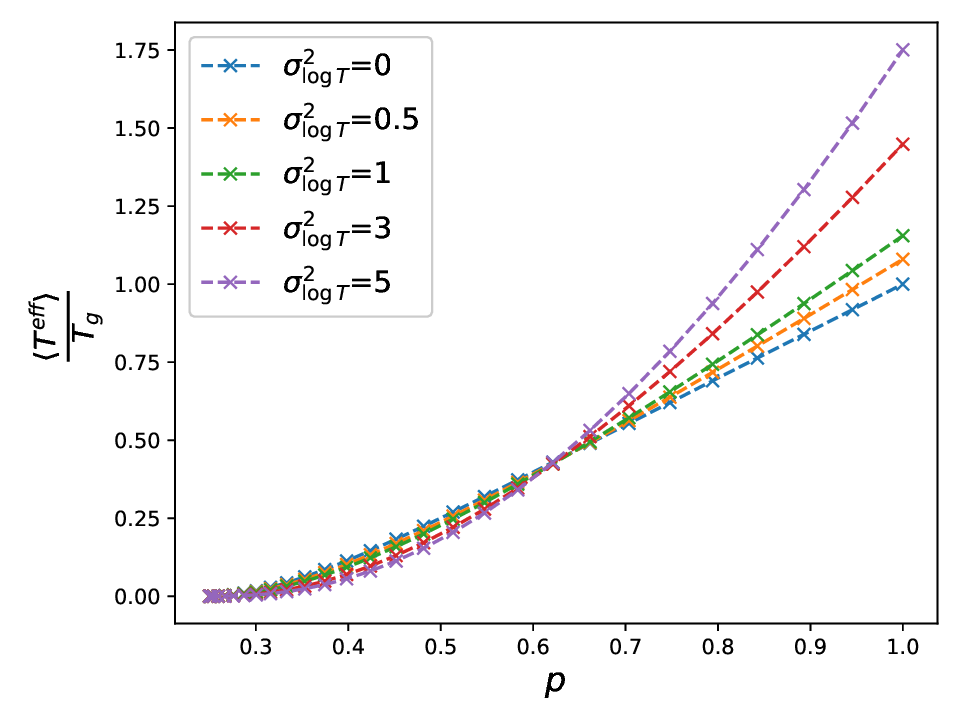}
\caption{Arithmetic mean of $T^{eff}$ for structured RRNs: (left) in 2D, and (right) in 3D. In 2D networks, $T^{eff}$ decreases uniformly with increasing $\sigma_{\log T}^2$ across all $p$ values. Conversely, in 3D networks, $T^{eff}$ decreases when $\sigma_{\log T}^2$ increases at $p$ values below 0.63, and increases for $p$ values above this threshold. Notably, at full connectivity ($p=1$) for 2D and $p=0.63$ for 3D, $T^{eff}$ remains stable despite changes in variance, demonstrating effective compensation by the network's geometry.}
   \label{fig:Keff}
\end{figure}

In this section, we analyze the behavior of the average of the effective conductivity over the $N_{real}$, $\langle T^{eff} \rangle$, as a function of the connectivity parameter $p$ across structured Random Resistor Networks (RRNs) with $r=1$. The effective conductivity is computed for each parameter combination, and the results are presented in Fig.~\ref{fig:Keff}.

In 2D networks, effective conductivity consistently decreases with increasing $\sigma_{\log T}^2$, regardless of $p$. In 3D networks, the response to increasing $\sigma_{\log T}^2$ is more nuanced: effective conductivity decreases when $p$ is less than approximately 0.63, but increases for larger $p$ values.

The increase in variance, $\sigma_{\log T}^2$\added{,} depending on the network's geometry, can lead to the formation of high-conductivity paths or to situations where the low conductivity of certain edges dominates the total flow. \replaced{Conversely}{In opposition}, at $p=1$ for 2D networks and at $p=0.63$ for 3D networks, $T^{eff}$ shows independence from the variance. This indicates a unique point of compensation where the network's geometry or connectivity counterbalances the effects of increased variance. 

\begin{figure}[h!]
  \centering
  \includegraphics[width=0.48\textwidth]{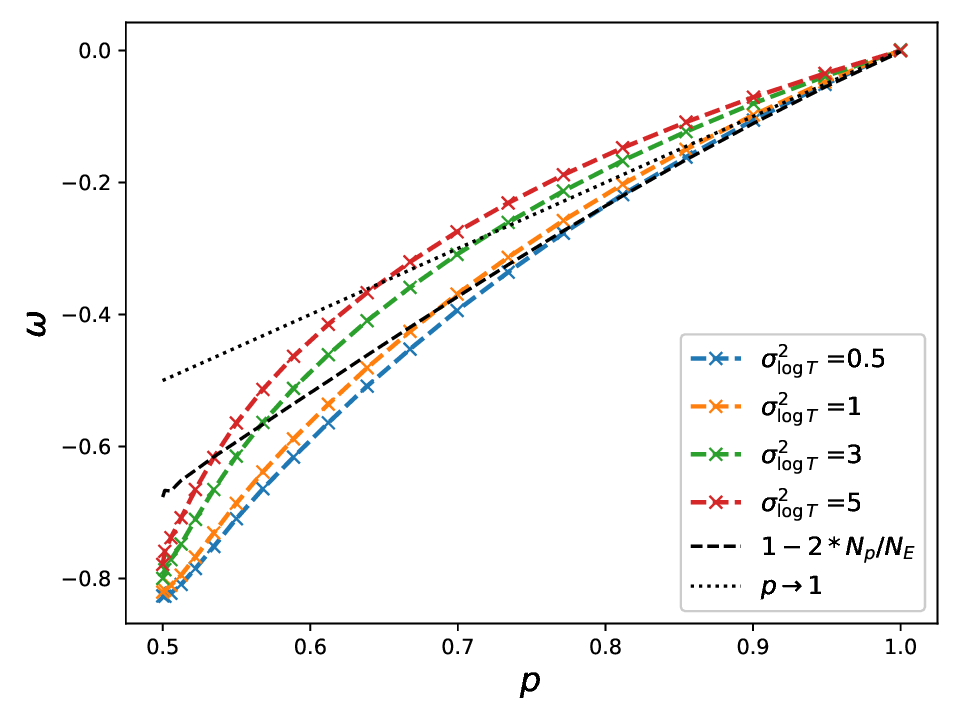}
  \includegraphics[width=0.48\textwidth]{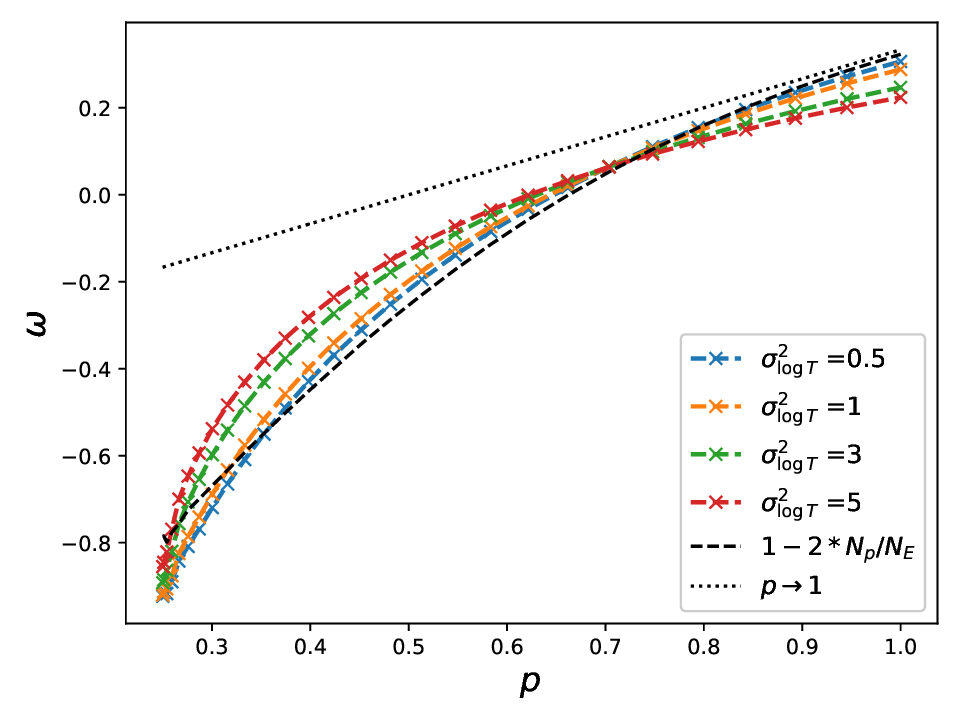}
   \caption{Power average exponent $\omega$, obtained from the results shown in Fig.~\ref{fig:Keff}, along with the approximation given by Eq.~\ref{eq:NsNb} (represented by the full black line). The results are presented for 2D (Left) and 3D (Right).}
   \label{fig:OmegaApp}
\end{figure}

The mean value of the power average exponent $\omega$, for a RRN with a given $\sigma_{\log T}^2$, computed from the numerical results of $T^{eff}$\added{,} \replaced{is}{are} shown in Fig.~\ref{fig:OmegaApp}.

At this stage, the exponent $\omega$ appears only as a particular data post-processing\added{,} although its value may be estimated from theory for $p=1$. In that case, it was shown by numerical simulations that Eq.~\ref{eq:powerAvLogn} provides a robust estimator of $T^{eff}$ even far from its theoretical validity\added{,} limited to small log-conductivity variance. In \replaced{the present case}{present case}, that treatment may be of interest if the exponent $\omega$ has little dependence \replaced{on}{with} $\sigma_{\log T}^2$ and if its dependence \replaced{on}{with} $p$ may be related to overall geometrical properties of the supporting connected network.

It can be observed that both sets of curves depend on $\sigma_{\log T}^2$\added{,} \replaced{with the notable exception of}{at the notable exception for} $D=2$, $p=1$\added{,} that corresponds to the exact result of Matheron (1967). In 3D, for $p$ close to 0.72, the apparent exponent $\omega$ vanishes, corresponding to an apparent flow dimension of 2. As the exponent $\omega$ depends on the variance, it means that the mean-field hypothesis breaks down, even if it provides quite good first approximations. As a preliminary conclusion, in spite of their simplicity, power-law averaging formulae can provide a fast determination of the average conductivity of a random network.

\FloatBarrier

\subsection{\label{subsec:poweraverage2D3DrGT1}Effective conductivity and power average exponent of 2D and 3D Structured RRN with $r>1$}

\begin{figure}[h!]
    \centering
    \includegraphics[width=0.48\textwidth]{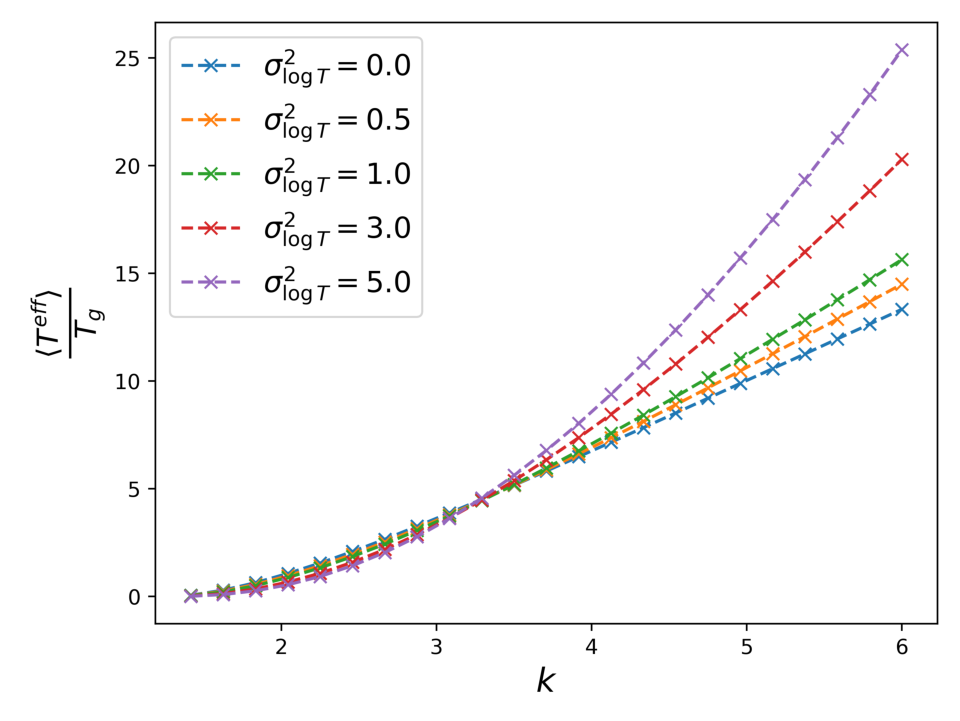}
    \includegraphics[width=0.48\textwidth]{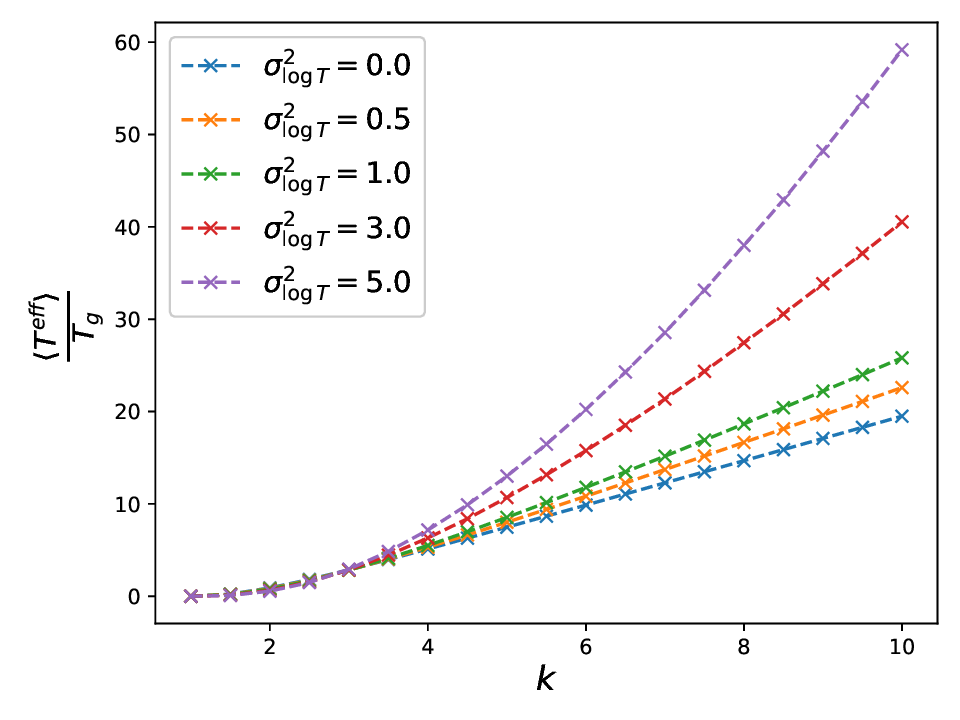}
    \caption{Variation of effective conductivity, $T^{eff}$, as a function of the mean degree $k$ for structured networks with a connectivity radius of $r=5$. The results are presented for 2D (left) and 3D (right).}
    \label{fig:Teff_radius}
\end{figure}

In Sec.~\ref{subsec:2D2Dresultsr=1l}, we analyzed the behavior of the effective conductivity $T^{eff}$ in RRNs with a connectivity radius $r$ of 1. Extending this analysis, Fig.~\ref{fig:Teff_radius} presents $T^{eff}$ of RRNs where the connectivity radius $r$ extends beyond immediate neighbors ($r=5$), as a function of the connectivity degree $k$.

In 2D networks, for values of $k$ close to 3.2, $T^{eff}$ shows independence from $\sigma_{\log T}^2$\added{,} analogous to the compensation point observed at $p=1$ ($k=4$) for $r=1$. As $k$ increases beyond this point, there is a pronounced increase in $T^{eff}$ with the variance.

For 3D networks, the compensation point similarly occurs around $k=3.2$. Beyond this threshold, $T^{eff}$ increases significantly faster with the variance for higher values of $k$ compared to 2D configurations. \added{Getting} a full understanding of the dependence of $\omega$ \replaced{on}{with} $p$ will be the topic of another study.

\begin{figure}[h!]
    \centering
    \includegraphics[width=0.48\textwidth]{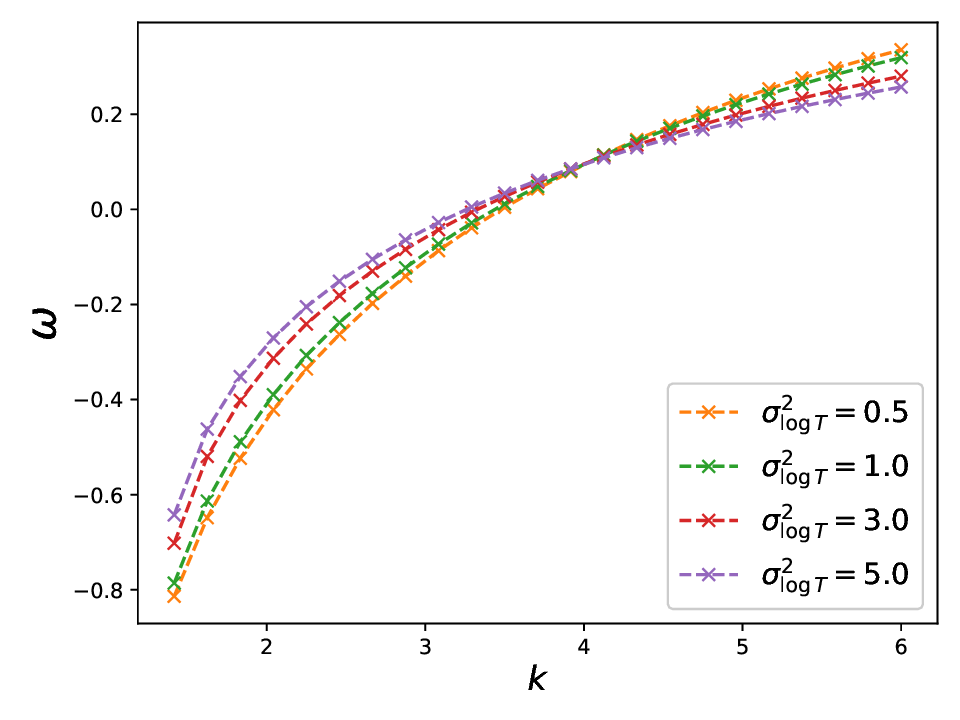}
    \includegraphics[width=0.48\textwidth]{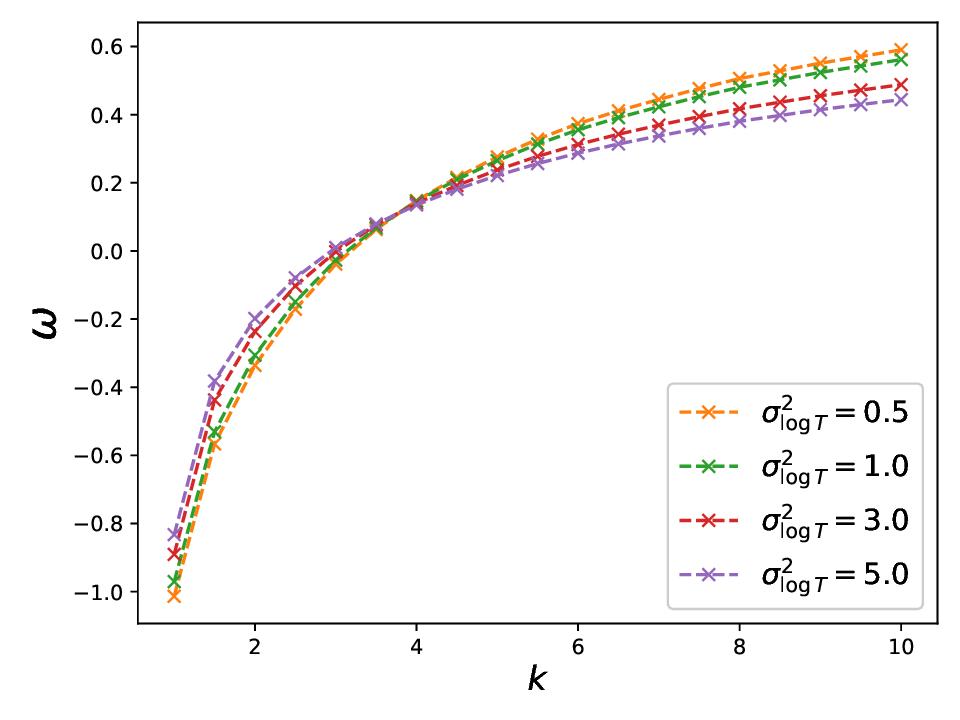}
    \caption{Power average exponent $\omega$ as a function of the mean degree $k$ for structured networks with a connectivity radius of $r=5$. The results, derived from the effective conductivity data shown in Fig.~\ref{fig:Teff_radius}, are presented for 2D networks on the left and 3D networks on the right.}
    \label{fig:omega_radius}
\end{figure}

The mean value of $\omega$, for a RRN with a given $\sigma_{\log T}^2$, is derived from the numerical results of $T^{eff}$ and displayed in Fig.~\ref{fig:omega_radius} for networks with a connectivity radius of $r=5$. The exponent $\omega$ maintains stability even in networks with enhanced connectivity. Notably, the compensation point where $\omega$ is independent of the variance \added{occurs} at $k=4$ for both 2D and 3D networks.

\begin{figure}[h!]
    \centering
    \includegraphics[width=0.48\textwidth]{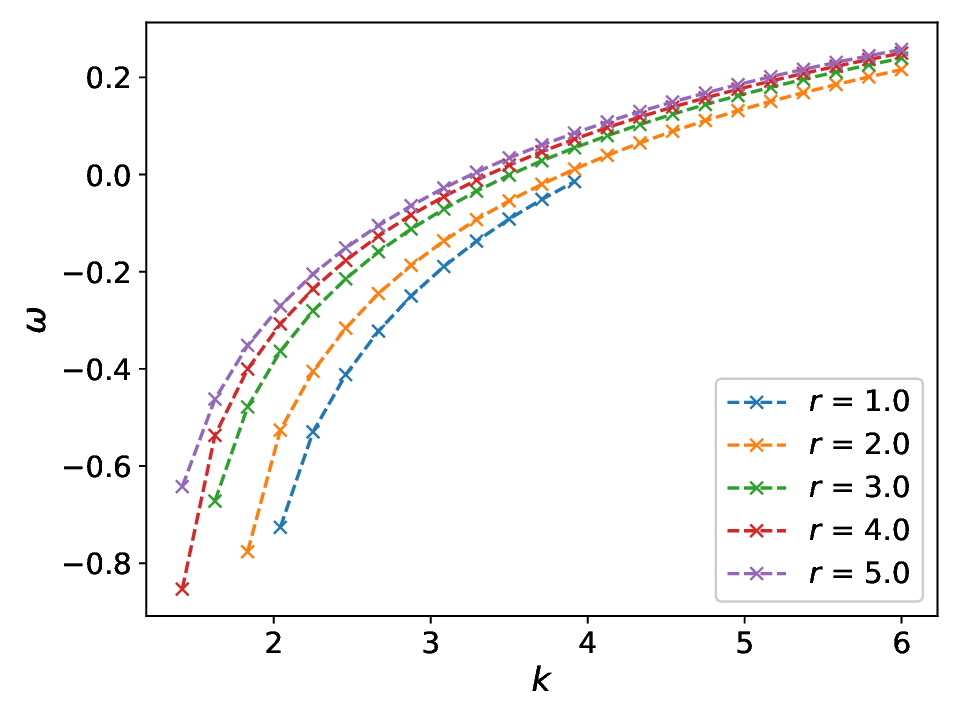}
    \includegraphics[width=0.48\textwidth]{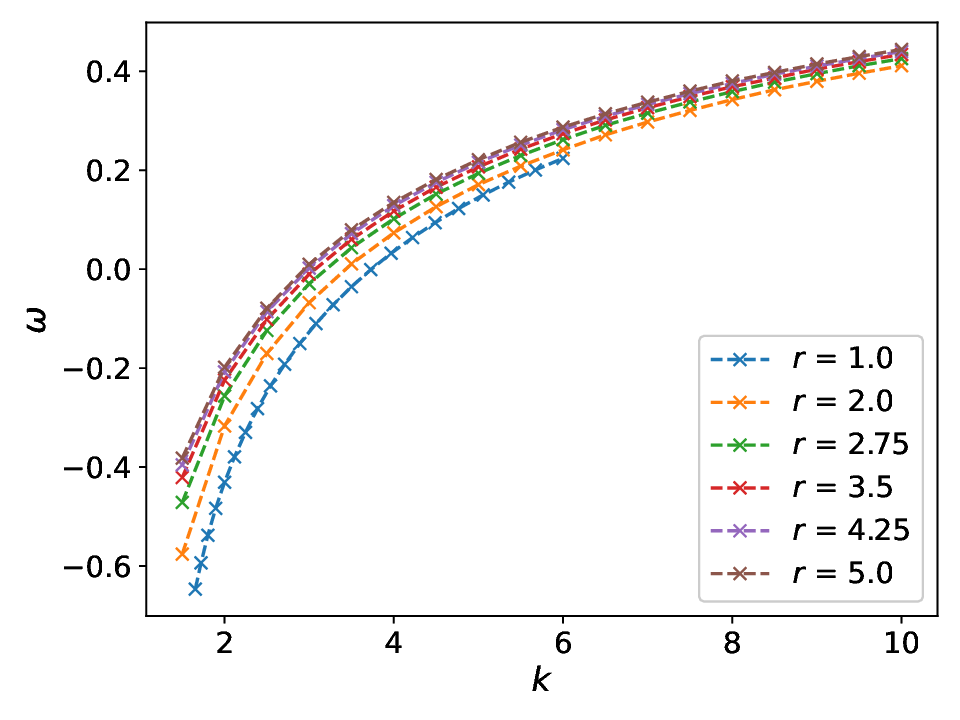}
    \caption{Variation of the power average exponent $\omega$ with the mean degree $k$ for structured networks, calculated for a high conductivity variance of $\sigma_{\log T}^2=5$. This set of graphs illustrates the influence of different connectivity radius on $\omega$ for 2D networks shown on the left and 3D networks on the right.}
    \label{fig:omega_k_var5}
\end{figure}

The results presented in Fig.~\ref{fig:omega_k_var5} illustrate the variation of the power average exponent $\omega$ with respect to the mean degree $k$ in structured networks for different values of $r$, under conditions of high conductivity variance ($\sigma_{\log T}^2=5$). These observations are consistent with percolation theory and stochastic modeling predictions, highlighting a decreased influence of $r$ as $k$ increases. This trend emphasizes the percolation transition, wherein the network transitions to a state of effective homogeneity at larger scales. Notably, the dimensionality of 3D networks results in a more expansive percolation threshold\added{,} consequently enhancing the network's overall transport capacity.

\FloatBarrier

\subsubsection{Effective Conductivity of RRN with $p=1$}
\label{subsub:full_net}

In this section, we present results for the effective conductivity $T^{eff}$ and the parameter $\omega$ for networks with $r > 1$, where no edges have been removed. This setup provides a baseline understanding of the system's behavior under maximal connectivity conditions. The difference between \replaced{these}{those} networks and Erdős–Rényi networks, which are characterized by edges placed randomly between nodes with a fixed probability, is that the underlying 2D or 3D structure imposes a natural way of computing effective parameters with well-defined inlet and outlet boundary conditions.

\begin{figure}[h!]
    \centering
    \includegraphics[width=0.48\textwidth]{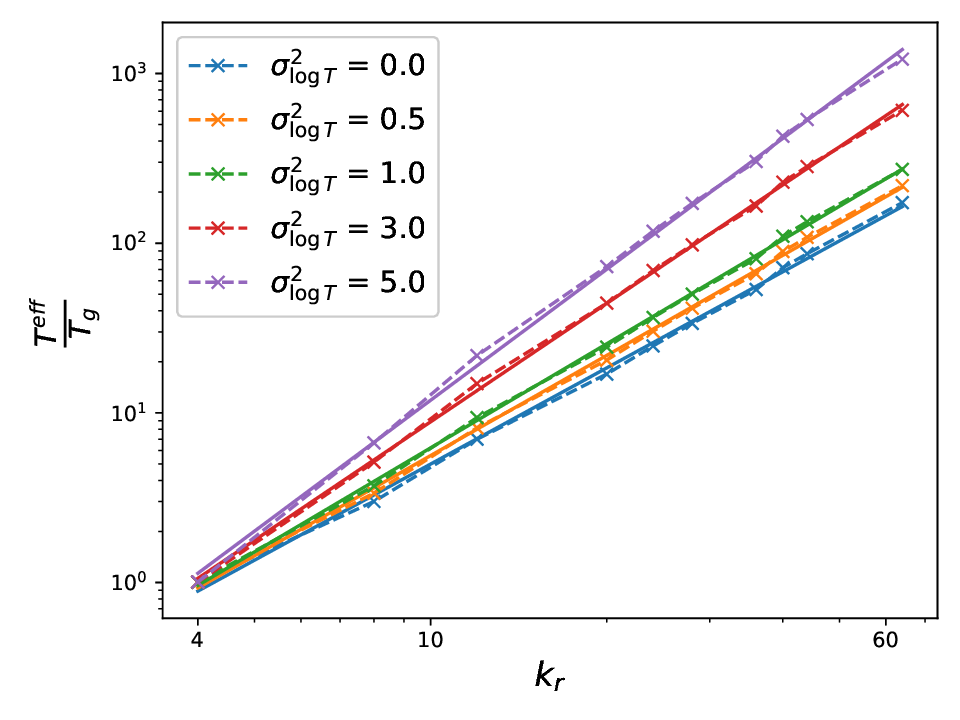}
    \includegraphics[width=0.48\textwidth]{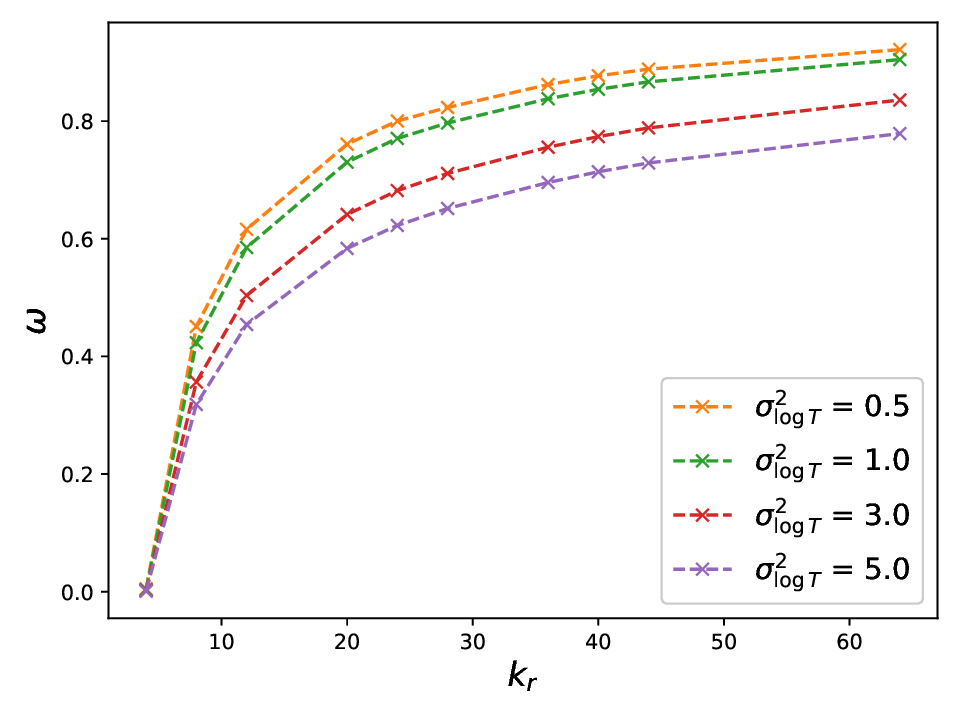}
    \caption{On the left, $T^{eff}$, of 2D RRN, normalized by the geometric mean $T_g$, is plotted against the mean degree $k$ on a logarithmic scale. The solid lines represent the fit to a power-law, with the power-law exponent listed in Table~\ref{tab:slopes}. On the right, the parameter $\omega$ is shown as a function of $k$, highlighting how network connectivity influences flow dynamics and transitions from localized to distributed flow regimes. These results establish a reference for the behavior of RRNs under maximum connectivity without edge removal.}         
    \label{exampleGnet_2D}
\end{figure}

The left panel of Fig.~\ref{exampleGnet_2D} depicts the variation of $T^{eff}$ as a function of the mean degree $k$, normalized by the geometric mean of the conductivity $T_g$. The relationship between $T^{eff}$ and $k$ suggests a power-law behavior, indicative of the network's scaling properties under full connectivity. The right panel shows the parameter $\omega$ across different values of $k$, illustrating how changes in network connectivity influence flow characteristics even without the introduction of disorder by edge removal.

For full networks with edges sharing the same conductivity, it can be \replaced{shown}{showed} that the overall conductivity scales with the square of its size. \replaced{This}{That} size may be replaced by $r$ in the case of large structured networks of connectivity radius $r$. \replaced{This is}{That is} mainly due to the number of connections growing as the square of the number of nodes. \replaced{This}{That} is confirmed in the left panel of Fig.~\ref{exampleGnet_2D} for $\sigma_{\log T}^2 = 0$.

Interestingly, $\omega$ shows a rapid increase from values close to zero at low $k$, corresponding to the classical Matheron's (1967) result for 2D networks, towards values approaching one, suggesting nearly parallel flow paths in the network as connectivity increases. This behavior highlights the transition from highly localized to more distributed flow as the network becomes increasingly connected, aligning with theoretical expectations for RRNs under varying connectivity and structural conditions.

The analysis of $T^{eff}$ and $\omega$ in fully connected networks sets the groundwork for understanding the impact of introducing disorder. It helps in distinguishing the effects of structural changes from those induced by altering connectivity and conductivity variance.

\begin{figure}[h!]
    \centering
    \includegraphics[width=0.48\textwidth]{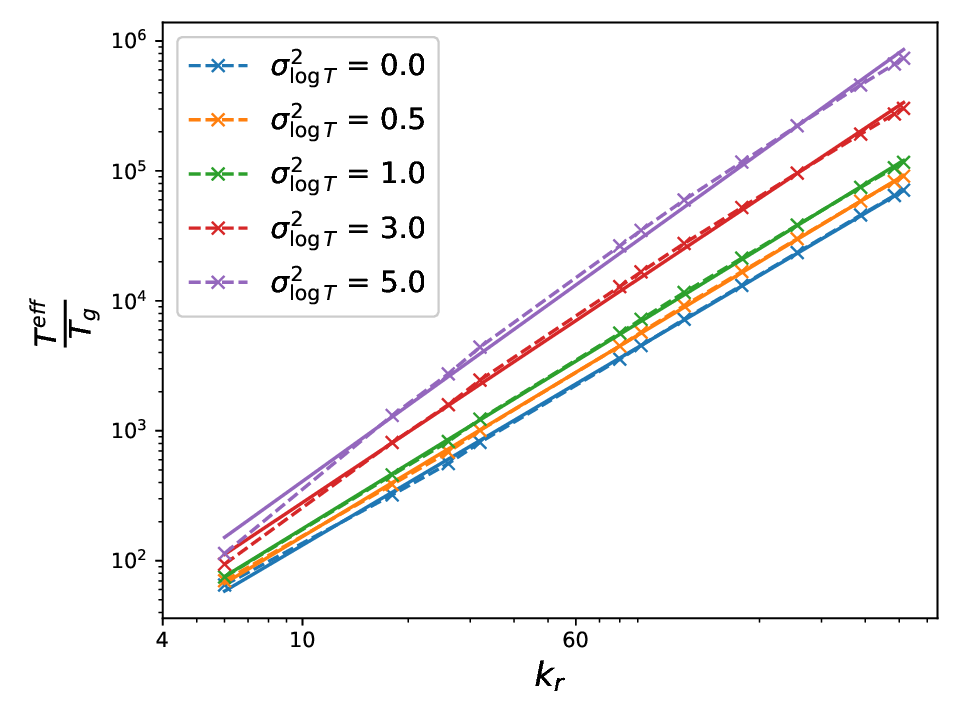}
    \includegraphics[width=0.48\textwidth]{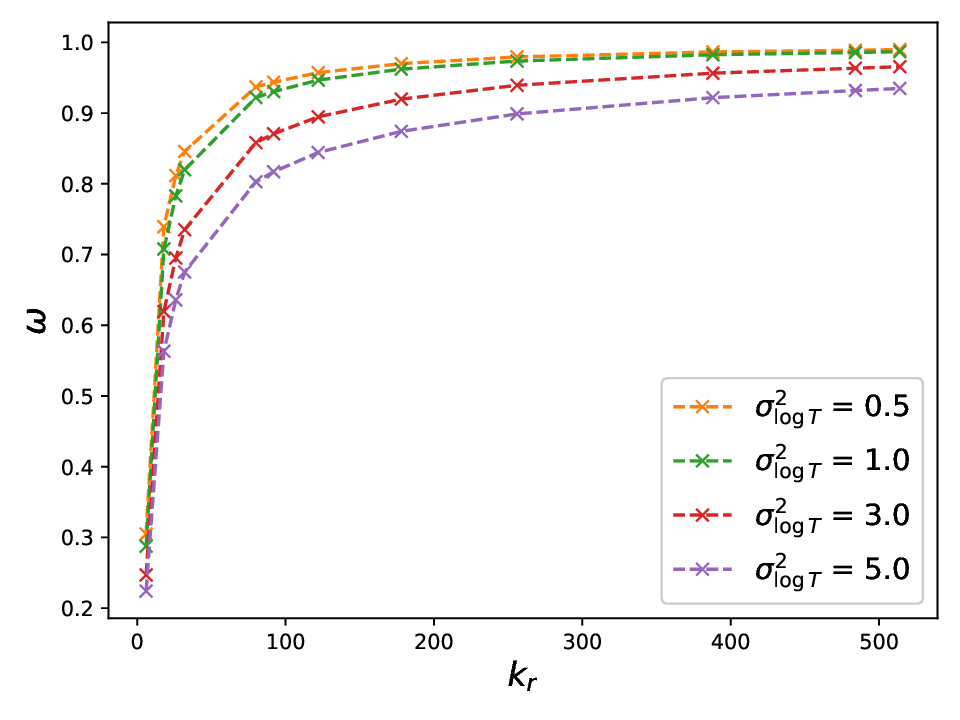}
    \caption{On the left, $T^{eff}$, of 3D RRN, normalized by the geometric mean conductivity $T_g$, is plotted against $k$ on a logarithmic scale. The solid lines represent the fit to a power-law, with the power-law exponent listed in Table~\ref{tab:slopes}. On the right, the parameter $\omega$ is shown as a function of $k$, highlighting how network connectivity influences flow dynamics and transitions from localized to distributed flow regimes.}         
    \label{exampleGnet3D}
\end{figure}
\begin{table}[h!]
\centering
\caption{Fitting Parameters for the Power Law Relationship $\frac{T^{eff}}{T_g} = C k_r^t$ for 2D and 3D Networks}
\label{tab:slopes}
\begin{tabular}{|c|c|c|c|c|c|}
\hline
\textbf{Dimension} & \multicolumn{5}{c|}{$\sigma^2_{\log T}$} \\
\cline{2-6}
 & \textbf{0.0} & \textbf{0.5} & \textbf{1.0} & \textbf{3.0} & \textbf{5.0} \\
\hline
$t$ (2D) & 1.88 & 1.96 & 2.04 & 2.32 & 2.56 \\
\hline
$t$ (3D) & 1.59 & 1.62 & 1.66 & 1.80 & 1.94 \\
\hline
\end{tabular}
\end{table}
This comprehensive analysis of $T^{eff}$ and $\omega$ under varying degrees of connectivity provides insights into the foundational dynamics of flow in RRNs. By comparing these results with those obtained from networks with introduced disorder (edge removal), we can better understand the intrinsic properties of RRNs and their response to structural perturbations.
\FloatBarrier

\section{Conclusions and discussion}\label{conc}

Results about the interplay between the randomness of edge conductivity and the geometry of the supporting network are reported. Using perturbation methods \replaced{that are}{quite} popular in the hydrology community, we \replaced{gain}{have} a better understanding of the effect of the local conductivity randomness \replaced{on}{about} the overall \replaced{behavior}{behaviour} of the RRN. The role of the so-called resistance distance is highlighted\added{, and} connections with up-to-date random graph/network theories can still lead to useful and original results. The conductivity randomness dominates for well-connected networks, while the topological disorder controls the overall \replaced{behavior}{behaviour} of the system close to the percolation threshold. The results can be summarized as follows:

\begin{itemize}
\item A procedure to average RRN's with respect to \replaced{fluctuations in the conductivities of their edges}{random conductivities fluctuations of the conductivity of its edges} was set up. The net result is an effective RRN with effective conductivities. Mathematically, it corresponds to estimating the harmonic average of the associated weighted Laplacian matrix of the RRN. The averaged operator keeps a Laplacian structure.

\item A suitable perturbation expansion was carried out. Strictly speaking, even with independent conductivities, the averaged RRN couples all the nodes, not just the input edges of the supporting RRN. However, up to the second order in a series expansion with respect to the logarithmic conductivity variance, the effective averaged RRN remains similar to the original one.

\item The averaged RRN is characterized by a set of local effective conductivities $T_{ij}^{eff}$ that depend on the set of so-called resistance distances introduced in graph theory. A mean-field theory \replaced{is used to}{allows to} propose a more robust estimation of the effective RRN using local power averaging of the local conductivities $T_{ij}$. The local averaging exponent depends on the geometry of the network.

\item A relation between the so-called overall effective conductivity of the RRN and the statistical properties of the conductivity was derived up to second order with respect to the log-conductivity variance, regardless \replaced{of}{on} the geometry of the network. A robust extrapolation under the form of a power-averaging formula was established. A mean-field approach was proposed to provide a more solid foundation for this claim in Appendix~\ref{App3}.

\item Numerical simulations \replaced{carried out}{carried-out} starting from Cartesian lattices in 2D or 3D are in correct agreement with \replaced{these}{that} theoretical predictions, although some discrepancy may be observed close to the percolation threshold.

\item For large networks, the results are independent \replaced{of}{on} the realization of both the network and the conductivity map, except close to the percolation threshold\added{,} as \replaced{expected}{it can be anticipated}. So\added{,} the proportion of active edges and the input conductivity distribution variances are the most relevant parameters.

\item The corresponding averaging exponent $\omega$ is quite stable \replaced{with respect to}{with} the input variance and may be related to the proportion $p$ of active edges.
\end{itemize}

\section{Perspectives}\label{perspec}
Several perspectives can be sketched:
\begin{itemize}
    \item On the theoretical side, the overall structure of the harmonic average of the weighted Laplacian should be studied for general log-conductivity variance in order to justify that its structure remains essentially similar to the structure of $L_0$, with weights decreasing exponentially with the distance between vertices. In that direction, in analogy with averaging continuous Laplace equations with random coefficients, spectral methods appear as \replaced{a}{being a} useful tool.

    \item In continuous upscaling theories of Darcy's law, a related approach involves analyzing the spectrum of the Laplacian matrix via Fourier transform techniques. Investigating the small frequency eigenvalues and eigenvectors of $\langle \mathbf{L}^{-1} \rangle ^{-1}$, which describe the system's overall behavior\added{,} may be a possible avenue of investigation.

    \item The mean-field approach that \replaced{is used to}{allows to} extrapolate a second-order expansion to the power law Eq.~\ref{Eq:powerave} must be tested numerically, as well as the approximations underlying the exponential solutions of Eqs.~\ref{Eq:diffequwij} and \ref{Eq:diffequwtrace}, by evaluating numerically the variations of the products $R_{ij} T_{ij}$ \replaced{as the}{as long as the} log-conductivity variance is increased.

    \item Performing a similar study using \replaced{non-structured}{as supporting network non-structured} RRNs, closer to networks encountered in practice\added{,} \replaced{is of interest, particularly in}{in particular in} geosciences applications.

    \item A related issue is \replaced{the selection of}{to select} various input conductivity distributions\added{;} indeed\added{,} choosing identically log-normally distributed independent conductivities is a strong hypothesis.

    \item We addressed the averaging issue that is closely related in the continuous case to the upscaling question (coarsening/homogenizing) of the flow equations. This is still relevant for RRN's\added{,} in which\added{,} \replaced{by}{in} analogy, the associated upscaling issue may be related to a suitable grouping of vertices and edges. Methods of graph sparsification relying on the resistance distance were developed by previous authors \citep{spielman2008graph} and may be useful for these upscaling/degree-of-freedom reduction purposes.

    \item Studying local flow, dissipation, and pressure drop distributions is essential for a deeper understanding of tracer transport, simplifying network models, and predicting the impact of specific edges on global flow\added{;} as well as addressing potential non-linearities in natural systems like karstic aquifers \citep{noetinger2013explicit}.
\end{itemize}

\begin{acknowledgments}
 The authors acknowledge funding by the European Union (ERC, KARST, 101071836). Views and opinions expressed are however those of the authors only and do not necessarily reflect those of the European Union or the European Research Council Executive Agency. Neither the European Union nor the granting authority can be held responsible for them.
\end{acknowledgments}

\appendix
\section{Mean-field argument}\label{App:meanfield}

We want to give some support to the power-law averaging extrapolated to higher log conductivity variance.
We introduce $T_{g ij}= exp \langle \log T_{ij} \rangle$ and $\sigma^2_{\log T_{ij}} \simeq \sigma^2_{ij}/\langle T_{ij}\rangle^2 $. Combining equations \ref{eq:Neuman2} and \ref{Eq:secondorder}, one gets, up to second order using the resistance distance $R_{ij}$:
\begin{equation}
  \langle \mathbf{L}^{-1} \rangle ^{-1}_{ij}=\langle T_{ij}\rangle - R_{ij} \sigma^2_{ij}\label{Eq:secondorderTij}
\end{equation}
So we get, in terms of effective RRN:
\begin{equation}
  T_{ij}^{eff}=\langle T_{ij}\rangle(1 -\langle T_{ij}\rangle R_{ij}\sigma^2_{\log T_{ij}})
\end{equation}
or, equivalently, up to second order:
\begin{equation}
  T_{ij}^{eff}=T_{g ij}+T_{g ij}(1/2 -\langle T_{ij}\rangle R_{ij})\sigma^2_{\log T_{ij}}
\end{equation}
\replaced{Since}{Because} up to second order, $\langle T_{ij}\rangle=\langle exp [\langle \log T_{ij} \rangle + \sigma_{\log T_{ij}} \zeta]\rangle \simeq T_{g ij}(1+1/2\sigma^2_{\log T_{ij}} +\dots)  $, where $\zeta$ is a random variable such that $\langle \zeta\rangle =0$ and $\langle \zeta^2\rangle =1$.

The resistance distance between i and j is \replaced{less than}{lower than} $1/T_{ij}$ because the current can flow through any path joining vertices i and j, including edge $ij$, so we get the following inequality: $0 \leq T_{ij}R_{ij}\leq 1$.\\

Now, the main idea is to add heterogeneity layer by layer by writing
\begin{equation}
T_{ij}=T_{g ij} exp (\sum_{K=1}^{M}\zeta_{K ij } \sigma_{\log T_{ij}}/\sqrt{M})
\end{equation}
where the $\zeta_{K ij }$ are independent random variables with respect to $K$, $i$ and $j$ such that $\langle \zeta_{K ij }\rangle =0$ and $\langle \zeta_{K ij }^2\rangle =1$.

\added{The} Central Limit Theorem states that when $M \rightarrow \infty$, the variable $\sum_{K=1}^{M}\zeta_{K ij } \sigma_{\log T_{ij}}/\sqrt{M}$ tends to a \replaced{Gaussian}{gaussian} variable of mean 0 and of variance $\sigma_{\log T_{ij}}^2$. \replaced{This}{That} implies that $T_{ij}$ is log-normal, corresponding to a product of a large number of positive random variables.

Then $T_{ij}^{eff,\ K}$ with $0\leq K \leq M$ is computed using equation \ref{Eq:secondorderTij} \replaced{which}{that} can be applied recursively, by averaging over $\zeta_{k ij }$ sequentially\added{,} and using the fact that $\sigma_{\log T_{ij}}^{2}/N$ is small, and replacing $\mathbf{L_{0}}$ by $\mathbf{L}^{(K-1)}$ built using the \replaced{previous $K-1$ heterogeneities}{K-1 previous heterogeneities}. We obtain:
\begin{equation}
T_{ij}^{eff,\ K}=T_{g ij}^{(K-1)}+T_{g ij}^{(K-1)}(1/2 -T_{g ij}^{(K-1)} R_{ij}^{(K-1)})\sigma_{\log T_{ij}}^2/M
\end{equation}
\replaced{Our main assumption is to substitute}{Our main hypothesis is to replace} $\langle T_{ij}^{(K-1)}\rangle $ by $T_{ij}^{eff,\ K-1}$. \replaced{This substitution implies}{That replacement means} that the K-th conductivity fluctuation interacts through the effective medium corresponding to the \replaced{preceding $K-1$ heterogeneity layers}{K-1 preceding heterogeneity fluctuations}. It corresponds to a mean-field approximation. So we obtain:
\begin{equation}
T_{ij}^{eff,\ K}=T_{ij}^{eff,\ K-1}+T_{ij}^{eff,\ K-1}(1/2 -T_{ij}^{eff,\ K-1} R_{ij}^{(K-1)})\sigma_{\log T_{ij}}^2/M
\end{equation}
\replaced{Letting}{Setting} $t=K/M$ and $dt=1/M$, \replaced{and taking the limit}{letting} $M \rightarrow \infty$, \replaced{this expression becomes}{that equation can be written under the form of} an ordinary differential equation given by:
\begin{equation}
\dfrac{d T_{ij}^{eff}(t) }{dt}=T_{ij}^{eff}(t)(1/2 -T_{ij}^{eff}(t) R_{ij}(t))\sigma_{\log T_{ij}}^2\label{Eq:diffequwij}
\end{equation}
\added{This equation is} to be integrated for $t\in [0,1]$ with initial condition $T_{ij}^{eff}(t=0) =T_{g ij}$.

\replaced{To proceed, we must estimate}{It remains to estimate} the variations of $T_{ij}^{eff}(t) R_{ij}(t)$. The simplest hypothesis is to \replaced{assume}{set} $T_{ij}^{eff}(t) R_{ij}(t) \rightarrow T_{g ij} R_{ij}(t=0)$. \replaced{This expression corresponds to}{It corresponds to} the product of the geometric mean of the local conductivity by the corresponding edge resistance, evaluated for the starting geometrically averaged network. \replaced{This quantity likely depends}{That quantity is likely to depend} mainly on the overall graph geometry and \replaced{exhibits}{to present} smooth variations. Such an assumption remains to be tested numerically.

Using that assumption, the differential equation becomes linear and may be integrated, the desired value being its value for $t=1$.
\begin{equation}
T_{ij}^{eff}=T_{g ij} \exp \left[(1/2 -T_{g ij} R_{ij}(t=0))\sigma_{\log T_{ij}}^2\right]
\end{equation}

For \added{a log-normally distributed} $T_{ij}$\replaced{, this expression can be}{log-normally distributed, it may be} written in the equivalent power-averaging form:
\begin{equation}
T_{ij}^{eff}=\langle T_{ij}^{(1 -2T_{g ij} R_{ij}(t=0))}\rangle ^{1/(1 -2T_{g ij} R_{ij}(t=0))}
\end{equation}

\replaced{This compact expression}{That compact form} can be used even if the input $T_{ij}$'s follow a different distribution, but with a more limited range of validity. \added{Note that} for an edge such that $T_{g ij} R_{ij}=1$, one obtains the harmonic average of the associated conductivity. This corresponds to an edge through which all the current flows when its vertices are connected to a battery, yielding a physically sound result. \replaced{The}{That} product of the conductivity $T_{g ij}$ and the resistance $R_{ij}$ is likely to exhibit smooth variations and to depend on the overall geometry of the supporting network.

\section{Derivation of effective conductivity Eq.~\ref{Eq:LLMD} and Eq.~\ref{eq:NsNb} up to second order and of its associated power average extrapolation} \label{App3} 

\replaced{In this appendix, rather than determining the effective conductivities of individual edges—which requires the costly computation of a complete set of $R_{ij}$ values by solving Laplace equations for each of the $N_E$ edges—we focus on estimating a single effective conductivity, denoted $\lambda_{eff}$, such that the Laplacian $\lambda_{eff} \mathbf{T_0}$ approximates the behavior of $\langle \mathbf{L}^{-1} \rangle^{-1}$.}{In this Appendix, instead of determining local edges effective conductivities, that require quite heavy determination of a whole set of $R_{ij}$'s implying solving the number of edges $N_E$ times Laplace equations that can be costly, the focus is given about determining a single effective conductivity called $\lambda_{eff}$ such that the laplacian $\lambda_{eff} \bf L_0$ mimics the action of $\langle \bf{L}^{-1} \rangle ^{-1}$.}

The simplest criterion is to identify the corresponding trace by writing
\begin{equation}
\text{Tr} [\lambda_{eff} \mathbf{L_{0}}]= \text{Tr} \langle \mathbf{L}^{-1} \rangle ^{-1}
\end{equation}
which implies that the trace of the effective Laplacian is conserved. Although this requirement may seem arbitrary, it can be heuristically justified by analogy with averaging flow in heterogeneous porous media or electrical conductivity, where the energy dissipation associated with a uniform pressure or potential gradient is given by $\simeq \sum_{edges <i,j>} L_{i j}= -Tr L$ due to the structure of Laplacian matrices. Thus, $\lambda_{eff}$ is given by:

\begin{equation}
\lambda_{eff}=\dfrac{\text{Tr} \langle \mathbf{L}^{-1} \rangle ^{-1}}{\text{Tr } \mathbf{L_{0}}}
\end{equation}

To evaluate how $\lambda_{eff}$ depends on the variance of the conductivity heterogeneity, we can apply reasoning similar to that presented in Appendix~\ref{App:meanfield}, using a mean-field approach. Second-order perturbation theory provides:
\begin{equation}
\text{Tr} \langle \mathbf{L}^{-1} \rangle ^{-1}{ij}= \text{Tr } \mathbf{L{0}} - \sum_{\text{edges}} R_{ij} \sigma^2_{ij}
\label{Eq:secondorderTraceij}
\end{equation}

We can thus employ the same strategy as in Appendix~\ref{App:meanfield}, adding randomness layer by layer and expanding up to second order using the previous layer as the average. We therefore introduce the following sequence:

\begin{equation}
\lambda_{eff}^{K}=\dfrac{\text{Tr} \langle \mathbf{L}^{K,\ -1} \rangle ^{-1}}{\text{Tr } \mathbf{L_{0}}}
\label{Eq:sequenceTraceij}
\end{equation}

\replaced{To establish the recursion, we write:}{In order to set-up the recursion, we can write:}
\begin{equation}
\lambda_{eff}^{K}=\dfrac{\text{Tr} \langle \mathbf{L}^{K,\ -1} \rangle ^{-1}}{\text{Tr } \langle \mathbf{L}^{K-1,\ -1} \rangle ^{-1}} \times \lambda_{eff}^{K-1}
\label{Eq:astucesequenceTraceij}
\end{equation}

The perturbation equation \ref{Eq:secondorderTraceij} can then be applied, replacing $\bf{L_{0}}$ by $\mathbf{L}^{K-1}$.

We add the following assumption: the log-conductivity variance $\sigma^2_{\log(T_{ij})} = \sigma^2_{\log(T)}$ is the same for all edges.

Using the framework developed in Appendix~\ref{App:meanfield}, and assuming the mean-field approximation, we arrive at the following differential equation:

\begin{equation}
\dfrac{d \lambda^{eff}(t)}{dt} = \lambda^{eff}(t)\left(\dfrac{1}{2} - \dfrac{\sum_{\text{edges } ij} \left[T_{g\ ij}^{eff}(t=0)\right]^2 R_{ij}(t)}{\sum_{\text{edges } ij} T^{eff}{g\ ij}}\right)\sigma{\log T}^2
\label{Eq:diffequwtrace}
\end{equation}

Then, the nonlinear term retains its initial value at $t = 0$, the equation becomes linear and can be integrated.

\begin{equation}
\dfrac{d \lambda ^{eff}(t) }{dt}=\lambda ^{eff}(t)(1/2 - \dfrac{\sum _{edges\ ij}T_{g ij}^{2}R_{ij}}{\sum _{edges\ ij}T_{g ij}})\sigma_{\log T}^2\label{Eq:diffequwtracet=0}
\end{equation}
Its solution is given by
\begin{equation}
\lambda ^{eff}(t)= exp (1/2 - \dfrac{\sum _{edges\ ij}T_{g ij}^{2}R_{ij}}{\sum _{edges\ ij}T_{g ij}})\sigma_{\log T}^2t\label{Eq:soldiffequwtracet=0}
\end{equation}
So finally at t=1.
\begin{equation}
\mathbf{L}^{eff}= \mathbf{L}_{g}exp (1/2 - \dfrac{\sum _{edges\ ij}T_{g ij}^{2}R_{ij}}{\sum _{edges\ ij}T_{g ij}})\sigma_{\log T}^2\label{Eq:effectiveTequwtracet=0}
\end{equation}
This justifies the relevance of power averaging in the log-normal case with an exponent $\omega$ given by
\begin{equation}
 \omega = 1 -2 \dfrac{\sum _{edges\ ij}T_{g ij}^{2}R_{ij}}{\sum _{edges\ ij}T_{g ij}}\label{Eq:omegageneral}
\end{equation}

This overall exponent\added{,} that appears to be a weighted average over the whole network of $T_{g ij}^{2}R_{ij}$\added{,} is likely to be more stable than its local counterpart of Appendix~\ref{App:meanfield}.\\
In order to estimate $\omega$ in the cases presented in Sec.~\ref{sec:resultsDdimensionalsructuredRRN}, \replaced{we consider}{considering} a large structured RRN of $N^D$ vertices in $D$ dimensions, corresponding to a finite difference approximation of a continuous Laplace equation using a $2D+1$ stencil. This corresponds to our lattice RRN with the parameter $r$ set to 1. 
\replaced{We neglect boundary effects}{Boundary effects are neglected} due to the large value of $N$.

We now proceed to compute
$\sum_{<i,j>}R_{ij}=\sum_{<i,j>} [{L_0}^{-1}_{ij}+{L_0}^{-1}_{ji}-{L_0}^{-1}_{ii}-{L_0}^{-1}_{jj}]$

That may be carried out by remarking $\mathbf{L_0}\mathbf{L_0}^{-1}=\mathbf{1}$; \added{using} the specific form of $\mathbf{L_0}$, that expression can be simplified. We obtain, using Eq.~\ref{Eq:omegageneral}, $\omega =1-2/D$ with Eq.~\ref{Eq:LLMD}.\\
A similar analysis can be carried out in the case of percolation networks, built from \replaced{previously full}{full preceding} lattices, in which a proportion $p$ of edges are kept, \replaced{thus modifying}{$\mathbf{L_0}$ accordingly, so changing} $\mathbf{L_0}$ accordingly. \replaced{By removing}{Removing} finite clusters and using again the relation $\mathbf{L_0}\mathbf{L_0}^{-1}=\mathbf{1}$, combined with the specific form of $\mathbf{L_{0}}$\added{,} one gets:
\begin{equation}
\omega \approx 1-2\frac{N_{p}}{N_{E}}
\label{eq:omegaApprox}
\end{equation}
\added{Note} that strictly speaking\added{,} such a formula is valid only at second order, even if using the mean-field argument appears to be a rather robust extrapolation. It is exact for 1D path graphs, and more generally for full $D$-dimensionallattices \replaced{where}{were} the result $\omega=1-2/D$ is recovered.

\section{Analytical solution for path graphs}\label{apppathgraphs}

Consider a path graph of $N_p$ vertices ordered from 1 to $N_p$, listed in the order $v_1$, $v_2$, \ldots, $v_{N_p}$, such that the edges are ($v_i$, $v_{i+1}$) where i = 1, 2, …, $N_p$ - 1. The set of equations of the RRN may be written as:
\begin{eqnarray}
(1-\delta_{i,1}) T_{i-1,i}(P_{i-1}-P_{i}) + (1-\delta_{i,N_p})T_{i,i+1}(P_{i+1}-P_{i}) = Q_{i}. \nonumber \\
\label{Eq:recursionpath}
\end{eqnarray}
The two Kronecker symbols $\delta_{.,.}$ for i=1 or $N_p$ account for the special case of the extreme vertices 1 and N. An elementary recursion \replaced{allows us}{permits} to rewrite the RRN equations in the equivalent form:

\begin{eqnarray}
\forall i=1, \cdots N_p-1, P_{i+1}-P_{i}= \frac{1}{T_{i,i+1}}\sum_{k=1}^{i} Q_k
\label{Eq:solpath}
\end{eqnarray}

Specializing this expression for $i = N_p - 1$, we get 
$P_{N_p}-P_{N_p-1}=\frac{1}{T_{N_{p}-1,N_p}}\sum_{k=1}^{N_{p}-1} Q_{k}= -\frac{1}{T_{N_{p}-1,N_p}} Q_{N_p}$,
because $\sum_{j} Q_{j} =0$. \replaced{This}{That} corresponds to the conservation relation Eq.~\ref{Eq:recursionpath} \replaced{evaluated}{written} at vertex $v_{N_p}$. The existence of a solution is thus ensured\added{, and it can be made explicit by imposing} the normalization condition $\sum_{j} P_{j} =0$.

Averaging Eq.~\ref{Eq:solpath} over the disorder of the $T_{i,j}$ \replaced{shows that the effective averaged RRN has the same structure as the original one, with conductivities replaced by harmonic averages}{shows that the effective average RRN shares the same form than the original one, by replacing the set of conductivities by an effective set, sharing values given by the harmonic averages} $\langle \frac{1}{T_{i,i+1}}\rangle ^{-1}$, \replaced{as expected}{as it could be anticipated}.\\

Returning to the second-order expansion, and considering source terms of the form $Q_k=Q( \delta_{kj}-\delta_{ki})$, \replaced{the earlier solution}{the preceding solution} Eq.~\ref{Eq:solpath} \replaced{reveals that}{shows that}

\begin{equation}
    [{L_0}^{-1}_{ij}+{L_0}^{-1}_{ji}-{L_0}^{-1}_{ii}-{L_0}^{-1}_{jj}]Q=(P_i -P_j)= -\frac{Q}{\langle T \rangle} |j-i|
\end{equation}
So 
$$[{L_0}^{-1}_{ij}+{L_0}^{-1}_{ji}-{L_0}^{-1}_{ii}-{L_0}^{-1}_{jj}]=1/\langle T\rangle$$ for $j \in <i>$. That shows the consistency of the second-order expansion with the exact result for path graphs. The associated mean-field result leading to harmonic power averaging appears to be exact.

\bibliography{aapmsamp}

\providecommand{\noopsort}[1]{}\providecommand{\singleletter}[1]{#1}%
\begin{thebibliography}{62}%
\makeatletter
\providecommand \@ifxundefined [1]{%
 \@ifx{#1\undefined}
}%
\providecommand \@ifnum [1]{%
 \ifnum #1\expandafter \@firstoftwo
 \else \expandafter \@secondoftwo
 \fi
}%
\providecommand \@ifx [1]{%
 \ifx #1\expandafter \@firstoftwo
 \else \expandafter \@secondoftwo
 \fi
}%
\providecommand \natexlab [1]{#1}%
\providecommand \enquote  [1]{``#1''}%
\providecommand \bibnamefont  [1]{#1}%
\providecommand \bibfnamefont [1]{#1}%
\providecommand \citenamefont [1]{#1}%
\providecommand \href@noop [0]{\@secondoftwo}%
\providecommand \href [0]{\begingroup \@sanitize@url \@href}%
\providecommand \@href[1]{\@@startlink{#1}\@@href}%
\providecommand \@@href[1]{\endgroup#1\@@endlink}%
\providecommand \@sanitize@url [0]{\catcode `\\12\catcode `\$12\catcode `\&12\catcode `\#12\catcode `\^12\catcode `\_12\catcode `\%12\relax}%
\providecommand \@@startlink[1]{}%
\providecommand \@@endlink[0]{}%
\providecommand \url  [0]{\begingroup\@sanitize@url \@url }%
\providecommand \@url [1]{\endgroup\@href {#1}{\urlprefix }}%
\providecommand \urlprefix  [0]{URL }%
\providecommand \Eprint [0]{\href }%
\providecommand \doibase [0]{https://doi.org/}%
\providecommand \selectlanguage [0]{\@gobble}%
\providecommand \bibinfo  [0]{\@secondoftwo}%
\providecommand \bibfield  [0]{\@secondoftwo}%
\providecommand \translation [1]{[#1]}%
\providecommand \BibitemOpen [0]{}%
\providecommand \bibitemStop [0]{}%
\providecommand \bibitemNoStop [0]{.\EOS\space}%
\providecommand \EOS [0]{\spacefactor3000\relax}%
\providecommand \BibitemShut  [1]{\csname bibitem#1\endcsname}%
\let\auto@bib@innerbib\@empty
\bibitem [{\citenamefont {Maxwell}(1873)}]{maxwell1873treatise}%
  \BibitemOpen
  \bibfield  {author} {\bibinfo {author} {\bibfnamefont {J.~C.}\ \bibnamefont {Maxwell}},\ }\href@noop {} {\emph {\bibinfo {title} {A treatise on electricity and magnetism}}},\ Vol.~\bibinfo {volume} {1}\ (\bibinfo  {publisher} {Oxford: Clarendon Press},\ \bibinfo {year} {1873})\BibitemShut {NoStop}%
\bibitem [{\citenamefont {Landau}\ and\ \citenamefont {Lifshitz}(1960)}]{landau1960lifshitz}%
  \BibitemOpen
  \bibfield  {author} {\bibinfo {author} {\bibfnamefont {L.}~\bibnamefont {Landau}}\ and\ \bibinfo {author} {\bibfnamefont {E.}~\bibnamefont {Lifshitz}},\ }\href@noop {} {\emph {\bibinfo {title} {Electrodynamics of continuous media}}},\ Vol.~\bibinfo {volume} {8}\ (\bibinfo  {publisher} {Pergamon Press},\ \bibinfo {year} {1960})\ p.~\bibinfo {pages} {15}\BibitemShut {NoStop}%
\bibitem [{\citenamefont {Hashin}\ and\ \citenamefont {Shtrikman}(1962)}]{hashin1962variational}%
  \BibitemOpen
  \bibfield  {author} {\bibinfo {author} {\bibfnamefont {Z.}~\bibnamefont {Hashin}}\ and\ \bibinfo {author} {\bibfnamefont {S.}~\bibnamefont {Shtrikman}},\ }\bibfield  {title} {\enquote {\bibinfo {title} {A variational approach to the theory of the effective magnetic permeability of multiphase materials},}\ }\href@noop {} {\bibfield  {journal} {\bibinfo  {journal} {Journal of applied Physics}\ }\textbf {\bibinfo {volume} {33}},\ \bibinfo {pages} {3125--3131} (\bibinfo {year} {1962})}\BibitemShut {NoStop}%
\bibitem [{\citenamefont {Matheron}(1967)}]{matheron1967}%
  \BibitemOpen
  \bibfield  {author} {\bibinfo {author} {\bibfnamefont {G.}~\bibnamefont {Matheron}},\ }\href@noop {} {\emph {\bibinfo {title} {El{\'e}ments pour une th{\'e}orie des milieux poreux}}}\ (\bibinfo  {publisher} {Masson},\ \bibinfo {year} {1967})\BibitemShut {NoStop}%
\bibitem [{\citenamefont {Torquato}\ and\ \citenamefont {Haslach~Jr}(2002)}]{torquato2002random}%
  \BibitemOpen
  \bibfield  {author} {\bibinfo {author} {\bibfnamefont {S.}~\bibnamefont {Torquato}}\ and\ \bibinfo {author} {\bibfnamefont {H.~W.}\ \bibnamefont {Haslach~Jr}},\ }\bibfield  {title} {\enquote {\bibinfo {title} {Random heterogeneous materials: microstructure and macroscopic properties},}\ }\href@noop {} {\bibfield  {journal} {\bibinfo  {journal} {Appl. Mech. Rev.}\ }\textbf {\bibinfo {volume} {55}},\ \bibinfo {pages} {B62--B63} (\bibinfo {year} {2002})}\BibitemShut {NoStop}%
\bibitem [{\citenamefont {Gelhar}(1993)}]{gelhar1993stochastic}%
  \BibitemOpen
  \bibfield  {author} {\bibinfo {author} {\bibfnamefont {L.~W.}\ \bibnamefont {Gelhar}},\ }\href@noop {} {\emph {\bibinfo {title} {Stochastic subsurface hydrology}}}\ (\bibinfo  {publisher} {Prentice-Hall},\ \bibinfo {year} {1993})\BibitemShut {NoStop}%
\bibitem [{\citenamefont {Dagan}(1989)}]{dagan1989flow}%
  \BibitemOpen
  \bibfield  {author} {\bibinfo {author} {\bibfnamefont {G.}~\bibnamefont {Dagan}},\ }\href@noop {} {\emph {\bibinfo {title} {Flow and transport in porous formations}}}\ (\bibinfo  {publisher} {Springer-Verlag GmbH \& Co. KG.},\ \bibinfo {year} {1989})\BibitemShut {NoStop}%
\bibitem [{\citenamefont {Indelman}\ and\ \citenamefont {Abramovich}(1994)}]{indelman1994higher}%
  \BibitemOpen
  \bibfield  {author} {\bibinfo {author} {\bibfnamefont {P.}~\bibnamefont {Indelman}}\ and\ \bibinfo {author} {\bibfnamefont {B.}~\bibnamefont {Abramovich}},\ }\bibfield  {title} {\enquote {\bibinfo {title} {A higher-order approximation to effective conductivity in media of anisotropic random structure},}\ }\href@noop {} {\bibfield  {journal} {\bibinfo  {journal} {Water resources research}\ }\textbf {\bibinfo {volume} {30}},\ \bibinfo {pages} {1857--1864} (\bibinfo {year} {1994})}\BibitemShut {NoStop}%
\bibitem [{\citenamefont {Abramovich}\ and\ \citenamefont {Indelman}(1995)}]{abramovich1995effective}%
  \BibitemOpen
  \bibfield  {author} {\bibinfo {author} {\bibfnamefont {B.}~\bibnamefont {Abramovich}}\ and\ \bibinfo {author} {\bibfnamefont {P.}~\bibnamefont {Indelman}},\ }\bibfield  {title} {\enquote {\bibinfo {title} {Effective permittivity of log-normal isotropic random media},}\ }\href@noop {} {\bibfield  {journal} {\bibinfo  {journal} {Journal of Physics A: Mathematical and General}\ }\textbf {\bibinfo {volume} {28}},\ \bibinfo {pages} {693} (\bibinfo {year} {1995})}\BibitemShut {NoStop}%
\bibitem [{\citenamefont {Renard}\ and\ \citenamefont {De~Marsily}(1997)}]{renard1997calculating}%
  \BibitemOpen
  \bibfield  {author} {\bibinfo {author} {\bibfnamefont {P.}~\bibnamefont {Renard}}\ and\ \bibinfo {author} {\bibfnamefont {G.}~\bibnamefont {De~Marsily}},\ }\bibfield  {title} {\enquote {\bibinfo {title} {Calculating equivalent permeability: a review},}\ }\href@noop {} {\bibfield  {journal} {\bibinfo  {journal} {Advances in water resources}\ }\textbf {\bibinfo {volume} {20}},\ \bibinfo {pages} {253--278} (\bibinfo {year} {1997})}\BibitemShut {NoStop}%
\bibitem [{\citenamefont {Jikov}, \citenamefont {Kozlov},\ and\ \citenamefont {Oleinik}(2012)}]{jikov2012homogenization}%
  \BibitemOpen
  \bibfield  {author} {\bibinfo {author} {\bibfnamefont {V.~V.}\ \bibnamefont {Jikov}}, \bibinfo {author} {\bibfnamefont {S.~M.}\ \bibnamefont {Kozlov}},\ and\ \bibinfo {author} {\bibfnamefont {O.~A.}\ \bibnamefont {Oleinik}},\ }\href@noop {} {\emph {\bibinfo {title} {Homogenization of differential operators and integral functionals}}}\ (\bibinfo  {publisher} {Springer Science \& Business Media},\ \bibinfo {year} {2012})\BibitemShut {NoStop}%
\bibitem [{\citenamefont {Armstrong}, \citenamefont {Kuusi},\ and\ \citenamefont {Mourrat}(2019)}]{armstrong2019quantitative}%
  \BibitemOpen
  \bibfield  {author} {\bibinfo {author} {\bibfnamefont {S.}~\bibnamefont {Armstrong}}, \bibinfo {author} {\bibfnamefont {T.}~\bibnamefont {Kuusi}},\ and\ \bibinfo {author} {\bibfnamefont {J.-C.}\ \bibnamefont {Mourrat}},\ }\href@noop {} {\emph {\bibinfo {title} {Quantitative stochastic homogenization and large-scale regularity}}},\ Vol.\ \bibinfo {volume} {352}\ (\bibinfo  {publisher} {Springer},\ \bibinfo {year} {2019})\BibitemShut {NoStop}%
\bibitem [{\citenamefont {King}(1987)}]{king1987use}%
  \BibitemOpen
  \bibfield  {author} {\bibinfo {author} {\bibfnamefont {P.}~\bibnamefont {King}},\ }\bibfield  {title} {\enquote {\bibinfo {title} {The use of field theoretic methods for the study of flow in a heterogeneous porous medium},}\ }\href@noop {} {\bibfield  {journal} {\bibinfo  {journal} {Journal of Physics A: Mathematical and General}\ }\textbf {\bibinfo {volume} {20}},\ \bibinfo {pages} {3935} (\bibinfo {year} {1987})}\BibitemShut {NoStop}%
\bibitem [{\citenamefont {Quintard}\ and\ \citenamefont {Whitaker}(1994)}]{quintard1994transport}%
  \BibitemOpen
  \bibfield  {author} {\bibinfo {author} {\bibfnamefont {M.}~\bibnamefont {Quintard}}\ and\ \bibinfo {author} {\bibfnamefont {S.}~\bibnamefont {Whitaker}},\ }\bibfield  {title} {\enquote {\bibinfo {title} {Transport in ordered and disordered porous media ii: Generalized volume averaging},}\ }\href@noop {} {\bibfield  {journal} {\bibinfo  {journal} {Transport in porous media}\ }\textbf {\bibinfo {volume} {14}},\ \bibinfo {pages} {179--206} (\bibinfo {year} {1994})}\BibitemShut {NoStop}%
\bibitem [{\citenamefont {Noetinger}(1994)}]{n1994effective}%
  \BibitemOpen
  \bibfield  {author} {\bibinfo {author} {\bibfnamefont {B.}~\bibnamefont {Noetinger}},\ }\bibfield  {title} {\enquote {\bibinfo {title} {The effective permeability of a heterogeneous porous medium},}\ }\href@noop {} {\bibfield  {journal} {\bibinfo  {journal} {Transport in porous media}\ }\textbf {\bibinfo {volume} {15}},\ \bibinfo {pages} {99--127} (\bibinfo {year} {1994})}\BibitemShut {NoStop}%
\bibitem [{\citenamefont {Noetinger}\ and\ \citenamefont {Gautier}(1998)}]{noetinger1998use}%
  \BibitemOpen
  \bibfield  {author} {\bibinfo {author} {\bibfnamefont {B.}~\bibnamefont {Noetinger}}\ and\ \bibinfo {author} {\bibfnamefont {Y.}~\bibnamefont {Gautier}},\ }\bibfield  {title} {\enquote {\bibinfo {title} {Use of the {F}ourier-{L}aplace transform and of diagrammatical methods to interpret pumping tests in heterogeneous reservoirs},}\ }\href@noop {} {\bibfield  {journal} {\bibinfo  {journal} {Advances in Water Resources}\ }\textbf {\bibinfo {volume} {21}},\ \bibinfo {pages} {581--590} (\bibinfo {year} {1998})}\BibitemShut {NoStop}%
\bibitem [{\citenamefont {Hristopulos}\ and\ \citenamefont {Christakos}(1999)}]{hristopulos1999renormalization}%
  \BibitemOpen
  \bibfield  {author} {\bibinfo {author} {\bibfnamefont {D.}~\bibnamefont {Hristopulos}}\ and\ \bibinfo {author} {\bibfnamefont {G.}~\bibnamefont {Christakos}},\ }\bibfield  {title} {\enquote {\bibinfo {title} {Renormalization group analysis of permeability upscaling},}\ }\href@noop {} {\bibfield  {journal} {\bibinfo  {journal} {Stochastic Environmental Research and Risk Assessment}\ }\textbf {\bibinfo {volume} {13}},\ \bibinfo {pages} {131--161} (\bibinfo {year} {1999})}\BibitemShut {NoStop}%
\bibitem [{\citenamefont {Nœtinger}(2000)}]{NOETINGER2000353}%
  \BibitemOpen
  \bibfield  {author} {\bibinfo {author} {\bibfnamefont {B.}~\bibnamefont {Nœtinger}},\ }\bibfield  {title} {\enquote {\bibinfo {title} {Computing the effective permeability of log-normal permeability fields using renormalization methods},}\ }\href {https://doi.org/https://doi.org/10.1016/S1251-8050(00)01412-9} {\bibfield  {journal} {\bibinfo  {journal} {Comptes Rendus de l'Académie des Sciences - Series IIA - Earth and Planetary Science}\ }\textbf {\bibinfo {volume} {331}},\ \bibinfo {pages} {353 -- 357} (\bibinfo {year} {2000})}\BibitemShut {NoStop}%
\bibitem [{\citenamefont {Teodorovich}(2002)}]{teodorovich2002renormalization}%
  \BibitemOpen
  \bibfield  {author} {\bibinfo {author} {\bibfnamefont {E.}~\bibnamefont {Teodorovich}},\ }\bibfield  {title} {\enquote {\bibinfo {title} {Renormalization group method in the problem of the effective conductivity of a randomly heterogeneous porous medium},}\ }\href@noop {} {\bibfield  {journal} {\bibinfo  {journal} {Journal of Experimental and Theoretical Physics}\ }\textbf {\bibinfo {volume} {95}},\ \bibinfo {pages} {67--76} (\bibinfo {year} {2002})}\BibitemShut {NoStop}%
\bibitem [{\citenamefont {Attinger}(2003)}]{attinger2003generalized}%
  \BibitemOpen
  \bibfield  {author} {\bibinfo {author} {\bibfnamefont {S.}~\bibnamefont {Attinger}},\ }\bibfield  {title} {\enquote {\bibinfo {title} {Generalized coarse graining procedures for flow in porous media},}\ }\href@noop {} {\bibfield  {journal} {\bibinfo  {journal} {Computational Geosciences}\ }\textbf {\bibinfo {volume} {7}},\ \bibinfo {pages} {253--273} (\bibinfo {year} {2003})}\BibitemShut {NoStop}%
\bibitem [{\citenamefont {Eberhard}, \citenamefont {Attinger},\ and\ \citenamefont {Wittum}(2004)}]{eberhard2004coarse}%
  \BibitemOpen
  \bibfield  {author} {\bibinfo {author} {\bibfnamefont {J.}~\bibnamefont {Eberhard}}, \bibinfo {author} {\bibfnamefont {S.}~\bibnamefont {Attinger}},\ and\ \bibinfo {author} {\bibfnamefont {G.}~\bibnamefont {Wittum}},\ }\bibfield  {title} {\enquote {\bibinfo {title} {Coarse graining for upscaling of flow in heterogeneous porous media},}\ }\href@noop {} {\bibfield  {journal} {\bibinfo  {journal} {Multiscale Modeling \& Simulation}\ }\textbf {\bibinfo {volume} {2}},\ \bibinfo {pages} {269--301} (\bibinfo {year} {2004})}\BibitemShut {NoStop}%
\bibitem [{\citenamefont {Stepanyants}\ and\ \citenamefont {Teodorovich}(2003)}]{stepanyants2003effective}%
  \BibitemOpen
  \bibfield  {author} {\bibinfo {author} {\bibfnamefont {Y.~A.}\ \bibnamefont {Stepanyants}}\ and\ \bibinfo {author} {\bibfnamefont {E.}~\bibnamefont {Teodorovich}},\ }\bibfield  {title} {\enquote {\bibinfo {title} {Effective hydraulic conductivity of a randomly heterogeneous porous medium},}\ }\href@noop {} {\bibfield  {journal} {\bibinfo  {journal} {Water Resources research}\ }\textbf {\bibinfo {volume} {39}} (\bibinfo {year} {2003})}\BibitemShut {NoStop}%
\bibitem [{\citenamefont {Hristopulos}(2020)}]{hristopulos2020random}%
  \BibitemOpen
  \bibfield  {author} {\bibinfo {author} {\bibfnamefont {D.~T.}\ \bibnamefont {Hristopulos}},\ }\href@noop {} {\emph {\bibinfo {title} {Random Fields for Spatial Data Modeling A Primer for Scientists and Engineers}}}\ (\bibinfo  {publisher} {Springer},\ \bibinfo {year} {2020})\BibitemShut {NoStop}%
\bibitem [{\citenamefont {Colecchio}\ \emph {et~al.}(2020)\citenamefont {Colecchio}, \citenamefont {Boschan}, \citenamefont {Otero},\ and\ \citenamefont {Noetinger}}]{colecchio2020multiscale}%
  \BibitemOpen
  \bibfield  {author} {\bibinfo {author} {\bibfnamefont {I.}~\bibnamefont {Colecchio}}, \bibinfo {author} {\bibfnamefont {A.}~\bibnamefont {Boschan}}, \bibinfo {author} {\bibfnamefont {A.~D.}\ \bibnamefont {Otero}},\ and\ \bibinfo {author} {\bibfnamefont {B.}~\bibnamefont {Noetinger}},\ }\bibfield  {title} {\enquote {\bibinfo {title} {On the multiscale characterization of effective hydraulic conductivity in random heterogeneous media: a historical survey and some new perspectives},}\ }\href@noop {} {\bibfield  {journal} {\bibinfo  {journal} {Advances in Water Resources}\ }\textbf {\bibinfo {volume} {140}},\ \bibinfo {pages} {103594} (\bibinfo {year} {2020})}\BibitemShut {NoStop}%
\bibitem [{\citenamefont {King}(1989)}]{king1989use}%
  \BibitemOpen
  \bibfield  {author} {\bibinfo {author} {\bibfnamefont {P.}~\bibnamefont {King}},\ }\bibfield  {title} {\enquote {\bibinfo {title} {The use of renormalization for calculating effective permeability},}\ }\href@noop {} {\bibfield  {journal} {\bibinfo  {journal} {Transport in porous media}\ }\textbf {\bibinfo {volume} {4}},\ \bibinfo {pages} {37--58} (\bibinfo {year} {1989})}\BibitemShut {NoStop}%
\bibitem [{\citenamefont {Berkowitz}\ and\ \citenamefont {Balberg}(1993)}]{berkowitz1993percolation}%
  \BibitemOpen
  \bibfield  {author} {\bibinfo {author} {\bibfnamefont {B.}~\bibnamefont {Berkowitz}}\ and\ \bibinfo {author} {\bibfnamefont {I.}~\bibnamefont {Balberg}},\ }\bibfield  {title} {\enquote {\bibinfo {title} {Percolation theory and its application to groundwater hydrology},}\ }\href@noop {} {\bibfield  {journal} {\bibinfo  {journal} {Water Resources Research}\ }\textbf {\bibinfo {volume} {29}},\ \bibinfo {pages} {775--794} (\bibinfo {year} {1993})}\BibitemShut {NoStop}%
\bibitem [{\citenamefont {Sahimi}(2011)}]{sahimi2011flow}%
  \BibitemOpen
  \bibfield  {author} {\bibinfo {author} {\bibfnamefont {M.}~\bibnamefont {Sahimi}},\ }\href@noop {} {\emph {\bibinfo {title} {Flow and transport in porous media and fractured rock: from classical methods to modern approaches}}}\ (\bibinfo  {publisher} {John Wiley \& Sons},\ \bibinfo {year} {2011})\BibitemShut {NoStop}%
\bibitem [{\citenamefont {Sahimi}(1994)}]{sahimi1994applications}%
  \BibitemOpen
  \bibfield  {author} {\bibinfo {author} {\bibfnamefont {M.}~\bibnamefont {Sahimi}},\ }\href@noop {} {\emph {\bibinfo {title} {Applications of percolation theory}}}\ (\bibinfo  {publisher} {CRC Press},\ \bibinfo {year} {1994})\BibitemShut {NoStop}%
\bibitem [{\citenamefont {Renard}\ and\ \citenamefont {Allard}(2013)}]{renard2013connectivity}%
  \BibitemOpen
  \bibfield  {author} {\bibinfo {author} {\bibfnamefont {P.}~\bibnamefont {Renard}}\ and\ \bibinfo {author} {\bibfnamefont {D.}~\bibnamefont {Allard}},\ }\bibfield  {title} {\enquote {\bibinfo {title} {Connectivity metrics for subsurface flow and transport},}\ }\href@noop {} {\bibfield  {journal} {\bibinfo  {journal} {Advances in Water Resources}\ }\textbf {\bibinfo {volume} {51}},\ \bibinfo {pages} {168--196} (\bibinfo {year} {2013})}\BibitemShut {NoStop}%
\bibitem [{\citenamefont {Hunt}, \citenamefont {Ewing},\ and\ \citenamefont {Ghanbarian}(2014)}]{hunt2014percolation}%
  \BibitemOpen
  \bibfield  {author} {\bibinfo {author} {\bibfnamefont {A.}~\bibnamefont {Hunt}}, \bibinfo {author} {\bibfnamefont {R.}~\bibnamefont {Ewing}},\ and\ \bibinfo {author} {\bibfnamefont {B.}~\bibnamefont {Ghanbarian}},\ }\href@noop {} {\emph {\bibinfo {title} {Percolation theory for flow in porous media}}},\ Vol.\ \bibinfo {volume} {880}\ (\bibinfo  {publisher} {Springer},\ \bibinfo {year} {2014})\BibitemShut {NoStop}%
\bibitem [{\citenamefont {Maillot}\ \emph {et~al.}(2016)\citenamefont {Maillot}, \citenamefont {Davy}, \citenamefont {Le~Goc}, \citenamefont {Darcel},\ and\ \citenamefont {De~Dreuzy}}]{maillot2016connectivity}%
  \BibitemOpen
  \bibfield  {author} {\bibinfo {author} {\bibfnamefont {J.}~\bibnamefont {Maillot}}, \bibinfo {author} {\bibfnamefont {P.}~\bibnamefont {Davy}}, \bibinfo {author} {\bibfnamefont {R.}~\bibnamefont {Le~Goc}}, \bibinfo {author} {\bibfnamefont {C.}~\bibnamefont {Darcel}},\ and\ \bibinfo {author} {\bibfnamefont {J.-R.}\ \bibnamefont {De~Dreuzy}},\ }\bibfield  {title} {\enquote {\bibinfo {title} {Connectivity, permeability, and channeling in randomly distributed and kinematically defined discrete fracture network models},}\ }\href@noop {} {\bibfield  {journal} {\bibinfo  {journal} {Water Resources Research}\ }\textbf {\bibinfo {volume} {52}},\ \bibinfo {pages} {8526--8545} (\bibinfo {year} {2016})}\BibitemShut {NoStop}%
\bibitem [{\citenamefont {Colecchio}\ \emph {et~al.}(2021)\citenamefont {Colecchio}, \citenamefont {Otero}, \citenamefont {Noetinger},\ and\ \citenamefont {Boschan}}]{colecchio2021equivalent}%
  \BibitemOpen
  \bibfield  {author} {\bibinfo {author} {\bibfnamefont {I.}~\bibnamefont {Colecchio}}, \bibinfo {author} {\bibfnamefont {A.~D.}\ \bibnamefont {Otero}}, \bibinfo {author} {\bibfnamefont {B.}~\bibnamefont {Noetinger}},\ and\ \bibinfo {author} {\bibfnamefont {A.}~\bibnamefont {Boschan}},\ }\bibfield  {title} {\enquote {\bibinfo {title} {Equivalent hydraulic conductivity, connectivity and percolation in 2d and 3d random binary media},}\ }\href@noop {} {\bibfield  {journal} {\bibinfo  {journal} {Advances in Water Resources}\ }\textbf {\bibinfo {volume} {158}},\ \bibinfo {pages} {104040} (\bibinfo {year} {2021})}\BibitemShut {NoStop}%
\bibitem [{\citenamefont {Halihan}\ and\ \citenamefont {Wicks}(1998)}]{HALIHAN1998}%
  \BibitemOpen
  \bibfield  {author} {\bibinfo {author} {\bibfnamefont {T.}~\bibnamefont {Halihan}}\ and\ \bibinfo {author} {\bibfnamefont {C.~M.}\ \bibnamefont {Wicks}},\ }\bibfield  {title} {\enquote {\bibinfo {title} {Modeling of storm responses in conduit flow aquifers with reservoirs},}\ }\href {https://doi.org/https://doi.org/10.1016/S0022-1694(98)00149-8} {\bibfield  {journal} {\bibinfo  {journal} {Journal of Hydrology}\ }\textbf {\bibinfo {volume} {208}},\ \bibinfo {pages} {82--91} (\bibinfo {year} {1998})}\BibitemShut {NoStop}%
\bibitem [{\citenamefont {Naughton}\ \emph {et~al.}(2017)\citenamefont {Naughton}, \citenamefont {Johnston}, \citenamefont {McCormack},\ and\ \citenamefont {Gill}}]{Naughton2017}%
  \BibitemOpen
  \bibfield  {author} {\bibinfo {author} {\bibfnamefont {O.}~\bibnamefont {Naughton}}, \bibinfo {author} {\bibfnamefont {P.}~\bibnamefont {Johnston}}, \bibinfo {author} {\bibfnamefont {T.}~\bibnamefont {McCormack}},\ and\ \bibinfo {author} {\bibfnamefont {L.}~\bibnamefont {Gill}},\ }\bibfield  {title} {\enquote {\bibinfo {title} {Groundwater flood risk mapping and management: examples from a lowland karst catchment in ireland},}\ }\href {https://doi.org/https://doi.org/10.1111/jfr3.12145} {\bibfield  {journal} {\bibinfo  {journal} {Journal of Flood Risk Management}\ }\textbf {\bibinfo {volume} {10}},\ \bibinfo {pages} {53--64} (\bibinfo {year} {2017})},\ \Eprint {https://arxiv.org/abs/https://onlinelibrary.wiley.com/doi/pdf/10.1111/jfr3.12145} {https://onlinelibrary.wiley.com/doi/pdf/10.1111/jfr3.12145} \BibitemShut {NoStop}%
\bibitem [{\citenamefont {Mayaud}\ \emph {et~al.}(2019)\citenamefont {Mayaud}, \citenamefont {Gabrovsek}, \citenamefont {Blatnik}, \citenamefont {Kogovšek}, \citenamefont {Petrič},\ and\ \citenamefont {Ravbar}}]{Mayaud2019}%
  \BibitemOpen
  \bibfield  {author} {\bibinfo {author} {\bibfnamefont {C.}~\bibnamefont {Mayaud}}, \bibinfo {author} {\bibfnamefont {F.}~\bibnamefont {Gabrovsek}}, \bibinfo {author} {\bibfnamefont {M.}~\bibnamefont {Blatnik}}, \bibinfo {author} {\bibfnamefont {B.}~\bibnamefont {Kogovšek}}, \bibinfo {author} {\bibfnamefont {M.}~\bibnamefont {Petrič}},\ and\ \bibinfo {author} {\bibfnamefont {N.}~\bibnamefont {Ravbar}},\ }\bibfield  {title} {\enquote {\bibinfo {title} {Understanding flooding in poljes: A modelling perspective},}\ }\href {https://doi.org/10.1016/j.jhydrol.2019.04.092} {\bibfield  {journal} {\bibinfo  {journal} {Journal of Hydrology}\ }\textbf {\bibinfo {volume} {575}},\ \bibinfo {pages} {874--889} (\bibinfo {year} {2019})}\BibitemShut {NoStop}%
\bibitem [{\citenamefont {Jeannin}\ \emph {et~al.}(2013)\citenamefont {Jeannin}, \citenamefont {Eichenberger}, \citenamefont {Sinreich} \emph {et~al.}}]{Jeannin2013}%
  \BibitemOpen
  \bibfield  {author} {\bibinfo {author} {\bibfnamefont {P.}~\bibnamefont {Jeannin}}, \bibinfo {author} {\bibfnamefont {U.}~\bibnamefont {Eichenberger}}, \bibinfo {author} {\bibfnamefont {M.}~\bibnamefont {Sinreich}}, \emph {et~al.},\ }\bibfield  {title} {\enquote {\bibinfo {title} {Karsys: a pragmatic approach to karst hydrogeological system conceptualisation. assessment of groundwater reserves and resources in switzerland},}\ }\href {https://doi.org/10.1007/s12665-012-1983-6} {\bibfield  {journal} {\bibinfo  {journal} {Environmental Earth Sciences}\ }\textbf {\bibinfo {volume} {69}},\ \bibinfo {pages} {999--1013} (\bibinfo {year} {2013})}\BibitemShut {NoStop}%
\bibitem [{\citenamefont {Jeannin}\ \emph {et~al.}(2021)\citenamefont {Jeannin}, \citenamefont {Artigue}, \citenamefont {Butscher}, \citenamefont {Chang}, \citenamefont {Charlier}, \citenamefont {Duran}, \citenamefont {Gill}, \citenamefont {Hartmann}, \citenamefont {Johannet}, \citenamefont {Jourde}, \citenamefont {Kavousi}, \citenamefont {Liesch}, \citenamefont {Liu}, \citenamefont {Lüthi}, \citenamefont {Malard}, \citenamefont {Mazzilli}, \citenamefont {Pardo-Igúzquiza}, \citenamefont {Thiéry}, \citenamefont {Reimann}, \citenamefont {Schuler}, \citenamefont {Wöhling},\ and\ \citenamefont {Wunsch}}]{JEANNIN2021126508}%
  \BibitemOpen
  \bibfield  {author} {\bibinfo {author} {\bibfnamefont {P.-Y.}\ \bibnamefont {Jeannin}}, \bibinfo {author} {\bibfnamefont {G.}~\bibnamefont {Artigue}}, \bibinfo {author} {\bibfnamefont {C.}~\bibnamefont {Butscher}}, \bibinfo {author} {\bibfnamefont {Y.}~\bibnamefont {Chang}}, \bibinfo {author} {\bibfnamefont {J.-B.}\ \bibnamefont {Charlier}}, \bibinfo {author} {\bibfnamefont {L.}~\bibnamefont {Duran}}, \bibinfo {author} {\bibfnamefont {L.}~\bibnamefont {Gill}}, \bibinfo {author} {\bibfnamefont {A.}~\bibnamefont {Hartmann}}, \bibinfo {author} {\bibfnamefont {A.}~\bibnamefont {Johannet}}, \bibinfo {author} {\bibfnamefont {H.}~\bibnamefont {Jourde}}, \bibinfo {author} {\bibfnamefont {A.}~\bibnamefont {Kavousi}}, \bibinfo {author} {\bibfnamefont {T.}~\bibnamefont {Liesch}}, \bibinfo {author} {\bibfnamefont {Y.}~\bibnamefont {Liu}}, \bibinfo {author} {\bibfnamefont {M.}~\bibnamefont {Lüthi}}, \bibinfo {author} {\bibfnamefont {A.}~\bibnamefont {Malard}}, \bibinfo {author} {\bibfnamefont {N.}~\bibnamefont
  {Mazzilli}}, \bibinfo {author} {\bibfnamefont {E.}~\bibnamefont {Pardo-Igúzquiza}}, \bibinfo {author} {\bibfnamefont {D.}~\bibnamefont {Thiéry}}, \bibinfo {author} {\bibfnamefont {T.}~\bibnamefont {Reimann}}, \bibinfo {author} {\bibfnamefont {P.}~\bibnamefont {Schuler}}, \bibinfo {author} {\bibfnamefont {T.}~\bibnamefont {Wöhling}},\ and\ \bibinfo {author} {\bibfnamefont {A.}~\bibnamefont {Wunsch}},\ }\bibfield  {title} {\enquote {\bibinfo {title} {Karst modelling challenge 1: Results of hydrological modelling},}\ }\href {https://doi.org/https://doi.org/10.1016/j.jhydrol.2021.126508} {\bibfield  {journal} {\bibinfo  {journal} {Journal of Hydrology}\ }\textbf {\bibinfo {volume} {600}},\ \bibinfo {pages} {126508} (\bibinfo {year} {2021})}\BibitemShut {NoStop}%
\bibitem [{\citenamefont {Gouy}\ \emph {et~al.}(2024)\citenamefont {Gouy}, \citenamefont {Collon}, \citenamefont {Bailly-Comte}, \citenamefont {Galin}, \citenamefont {Antoine}, \citenamefont {Thebault},\ and\ \citenamefont {Landrein}}]{GOUY2024130878}%
  \BibitemOpen
  \bibfield  {author} {\bibinfo {author} {\bibfnamefont {A.}~\bibnamefont {Gouy}}, \bibinfo {author} {\bibfnamefont {P.}~\bibnamefont {Collon}}, \bibinfo {author} {\bibfnamefont {V.}~\bibnamefont {Bailly-Comte}}, \bibinfo {author} {\bibfnamefont {E.}~\bibnamefont {Galin}}, \bibinfo {author} {\bibfnamefont {C.}~\bibnamefont {Antoine}}, \bibinfo {author} {\bibfnamefont {B.}~\bibnamefont {Thebault}},\ and\ \bibinfo {author} {\bibfnamefont {P.}~\bibnamefont {Landrein}},\ }\bibfield  {title} {\enquote {\bibinfo {title} {Karstnsim: A graph-based method for 3d geologically-driven simulation of karst networks},}\ }\href {https://doi.org/https://doi.org/10.1016/j.jhydrol.2024.130878} {\bibfield  {journal} {\bibinfo  {journal} {Journal of Hydrology}\ }\textbf {\bibinfo {volume} {632}},\ \bibinfo {pages} {130878} (\bibinfo {year} {2024})}\BibitemShut {NoStop}%
\bibitem [{\citenamefont {Frantz}\ \emph {et~al.}(2021)\citenamefont {Frantz}, \citenamefont {Collon}, \citenamefont {Renard},\ and\ \citenamefont {Viseur}}]{frantz2021}%
  \BibitemOpen
  \bibfield  {author} {\bibinfo {author} {\bibfnamefont {Y.}~\bibnamefont {Frantz}}, \bibinfo {author} {\bibfnamefont {P.}~\bibnamefont {Collon}}, \bibinfo {author} {\bibfnamefont {P.}~\bibnamefont {Renard}},\ and\ \bibinfo {author} {\bibfnamefont {S.}~\bibnamefont {Viseur}},\ }\bibfield  {title} {\enquote {\bibinfo {title} {Analysis and stochastic simulation of geometrical properties of conduits in karstic networks},}\ }\href {https://doi.org/https://doi.org/10.1016/j.geomorph.2020.107480} {\bibfield  {journal} {\bibinfo  {journal} {Geomorphology}\ }\textbf {\bibinfo {volume} {377}},\ \bibinfo {pages} {107480} (\bibinfo {year} {2021})}\BibitemShut {NoStop}%
\bibitem [{\citenamefont {Adler}, \citenamefont {Thovert},\ and\ \citenamefont {Mourzenko}(2013)}]{Adler20131}%
  \BibitemOpen
  \bibfield  {author} {\bibinfo {author} {\bibfnamefont {P.~M.}\ \bibnamefont {Adler}}, \bibinfo {author} {\bibfnamefont {J.-F.}\ \bibnamefont {Thovert}},\ and\ \bibinfo {author} {\bibfnamefont {V.~V.}\ \bibnamefont {Mourzenko}},\ }\href {https://doi.org/10.1093/acprof:oso/9780199666515.001.0001} {\emph {\bibinfo {title} {Fractured Porous Media}}},\ Vol.\ \bibinfo {volume} {9780199666515}\ (\bibinfo  {publisher} {Oxford University Press},\ \bibinfo {year} {2013})\ p.\ \bibinfo {pages} {1 – 184}\BibitemShut {NoStop}%
\bibitem [{\citenamefont {de~Dreuzy}, \citenamefont {Méheust},\ and\ \citenamefont {Pichot}(2012)}]{deDreuzy2012}%
  \BibitemOpen
  \bibfield  {author} {\bibinfo {author} {\bibfnamefont {J.-R.}\ \bibnamefont {de~Dreuzy}}, \bibinfo {author} {\bibfnamefont {Y.}~\bibnamefont {Méheust}},\ and\ \bibinfo {author} {\bibfnamefont {G.}~\bibnamefont {Pichot}},\ }\bibfield  {title} {\enquote {\bibinfo {title} {Influence of fracture scale heterogeneity on the flow properties of three-dimensional discrete fracture networks (dfn)},}\ }\href {https://doi.org/https://doi.org/10.1029/2012JB009461} {\bibfield  {journal} {\bibinfo  {journal} {Journal of Geophysical Research: Solid Earth}\ }\textbf {\bibinfo {volume} {117}} (\bibinfo {year} {2012}),\ https://doi.org/10.1029/2012JB009461},\ \Eprint {https://arxiv.org/abs/https://agupubs.onlinelibrary.wiley.com/doi/pdf/10.1029/2012JB009461} {https://agupubs.onlinelibrary.wiley.com/doi/pdf/10.1029/2012JB009461} \BibitemShut {NoStop}%
\bibitem [{\citenamefont {N{\oe}tinger}\ and\ \citenamefont {Jarrige}(2012)}]{noetinger2012quasi}%
  \BibitemOpen
  \bibfield  {author} {\bibinfo {author} {\bibfnamefont {B.}~\bibnamefont {N{\oe}tinger}}\ and\ \bibinfo {author} {\bibfnamefont {N.}~\bibnamefont {Jarrige}},\ }\bibfield  {title} {\enquote {\bibinfo {title} {A quasi steady state method for solving transient darcy flow in complex 3d fractured networks},}\ }\href@noop {} {\bibfield  {journal} {\bibinfo  {journal} {Journal of Computational Physics}\ }\textbf {\bibinfo {volume} {231}},\ \bibinfo {pages} {23--38} (\bibinfo {year} {2012})}\BibitemShut {NoStop}%
\bibitem [{\citenamefont {Hyman}\ \emph {et~al.}(2019)\citenamefont {Hyman}, \citenamefont {Dentz}, \citenamefont {Hagberg},\ and\ \citenamefont {Kang}}]{Hyman2019}%
  \BibitemOpen
  \bibfield  {author} {\bibinfo {author} {\bibfnamefont {J.~D.}\ \bibnamefont {Hyman}}, \bibinfo {author} {\bibfnamefont {M.}~\bibnamefont {Dentz}}, \bibinfo {author} {\bibfnamefont {A.}~\bibnamefont {Hagberg}},\ and\ \bibinfo {author} {\bibfnamefont {P.~K.}\ \bibnamefont {Kang}},\ }\bibfield  {title} {\enquote {\bibinfo {title} {Linking structural and transport properties in three-dimensional fracture networks},}\ }\href {https://doi.org/https://doi.org/10.1029/2018JB016553} {\bibfield  {journal} {\bibinfo  {journal} {Journal of Geophysical Research: Solid Earth}\ }\textbf {\bibinfo {volume} {124}},\ \bibinfo {pages} {1185--1204} (\bibinfo {year} {2019})},\ \Eprint {https://arxiv.org/abs/https://agupubs.onlinelibrary.wiley.com/doi/pdf/10.1029/2018JB016553} {https://agupubs.onlinelibrary.wiley.com/doi/pdf/10.1029/2018JB016553} \BibitemShut {NoStop}%
\bibitem [{\citenamefont {Noetinger}(2023)}]{noetinger2023random}%
  \BibitemOpen
  \bibfield  {author} {\bibinfo {author} {\bibfnamefont {B.}~\bibnamefont {Noetinger}},\ }\bibfield  {title} {\enquote {\bibinfo {title} {Random fields and up scaling, towards a more predictive probabilistic quantitative hydrogeology},}\ }\href@noop {} {\bibfield  {journal} {\bibinfo  {journal} {Comptes Rendus. G{\'e}oscience}\ }\textbf {\bibinfo {volume} {355}},\ \bibinfo {pages} {559--572} (\bibinfo {year} {2023})}\BibitemShut {NoStop}%
\bibitem [{\citenamefont {Journel}\ \emph {et~al.}(1986)\citenamefont {Journel}, \citenamefont {Deutsch}, \citenamefont {Desbarats} \emph {et~al.}}]{journel1986power}%
  \BibitemOpen
  \bibfield  {author} {\bibinfo {author} {\bibfnamefont {A.}~\bibnamefont {Journel}}, \bibinfo {author} {\bibfnamefont {C.}~\bibnamefont {Deutsch}}, \bibinfo {author} {\bibfnamefont {A.}~\bibnamefont {Desbarats}}, \emph {et~al.},\ }\bibfield  {title} {\enquote {\bibinfo {title} {Power averaging for block effective permeability},}\ }in\ \href@noop {} {\emph {\bibinfo {booktitle} {SPE California Regional Meeting}}}\ (\bibinfo {organization} {Society of Petroleum Engineers},\ \bibinfo {year} {1986})\BibitemShut {NoStop}%
\bibitem [{\citenamefont {Desbarats}(1992)}]{desbarats1992spatial}%
  \BibitemOpen
  \bibfield  {author} {\bibinfo {author} {\bibfnamefont {A.}~\bibnamefont {Desbarats}},\ }\bibfield  {title} {\enquote {\bibinfo {title} {Spatial averaging of hydraulic conductivity in three-dimensional heterogeneous porous media},}\ }\href@noop {} {\bibfield  {journal} {\bibinfo  {journal} {Mathematical Geology}\ }\textbf {\bibinfo {volume} {24}},\ \bibinfo {pages} {249--267} (\bibinfo {year} {1992})}\BibitemShut {NoStop}%
\bibitem [{\citenamefont {Neuman}\ and\ \citenamefont {Orr}(1993)}]{neuman1993prediction}%
  \BibitemOpen
  \bibfield  {author} {\bibinfo {author} {\bibfnamefont {S.~P.}\ \bibnamefont {Neuman}}\ and\ \bibinfo {author} {\bibfnamefont {S.}~\bibnamefont {Orr}},\ }\bibfield  {title} {\enquote {\bibinfo {title} {Prediction of steady state flow in nonuniform geologic media by conditional moments: Exact nonlocal formalism, effective conductivities, and weak approximation},}\ }\href@noop {} {\bibfield  {journal} {\bibinfo  {journal} {Water resources research}\ }\textbf {\bibinfo {volume} {29}},\ \bibinfo {pages} {341--364} (\bibinfo {year} {1993})}\BibitemShut {NoStop}%
\bibitem [{\citenamefont {Charlaix}, \citenamefont {Guyon},\ and\ \citenamefont {Roux}(1987)}]{Charlaix1987}%
  \BibitemOpen
  \bibfield  {author} {\bibinfo {author} {\bibfnamefont {E.}~\bibnamefont {Charlaix}}, \bibinfo {author} {\bibfnamefont {E.}~\bibnamefont {Guyon}},\ and\ \bibinfo {author} {\bibfnamefont {S.}~\bibnamefont {Roux}},\ }\bibfield  {title} {\enquote {\bibinfo {title} {Permeability of a random array of fractures of widely varying apertures},}\ }\href {https://doi.org/10.1007/BF00208535} {\bibfield  {journal} {\bibinfo  {journal} {Transport in Porous Media}\ }\textbf {\bibinfo {volume} {2}},\ \bibinfo {pages} {31--43} (\bibinfo {year} {1987})}\BibitemShut {NoStop}%
\bibitem [{\citenamefont {de~Dreuzy}\ \emph {et~al.}(2010)\citenamefont {de~Dreuzy}, \citenamefont {de~Boiry}, \citenamefont {Pichot},\ and\ \citenamefont {Davy}}]{deDreuzy2010}%
  \BibitemOpen
  \bibfield  {author} {\bibinfo {author} {\bibfnamefont {J.-R.}\ \bibnamefont {de~Dreuzy}}, \bibinfo {author} {\bibfnamefont {P.}~\bibnamefont {de~Boiry}}, \bibinfo {author} {\bibfnamefont {G.}~\bibnamefont {Pichot}},\ and\ \bibinfo {author} {\bibfnamefont {P.}~\bibnamefont {Davy}},\ }\bibfield  {title} {\enquote {\bibinfo {title} {Use of power averaging for quantifying the influence of structure organization on permeability upscaling in on-lattice networks under mean parallel flow},}\ }\href {https://doi.org/https://doi.org/10.1029/2009WR008769} {\bibfield  {journal} {\bibinfo  {journal} {Water Resources Research}\ }\textbf {\bibinfo {volume} {46}} (\bibinfo {year} {2010}),\ https://doi.org/10.1029/2009WR008769},\ \Eprint {https://arxiv.org/abs/https://agupubs.onlinelibrary.wiley.com/doi/pdf/10.1029/2009WR008769} {https://agupubs.onlinelibrary.wiley.com/doi/pdf/10.1029/2009WR008769} \BibitemShut {NoStop}%
\bibitem [{\citenamefont {Sanchez-Vila}, \citenamefont {Guadagnini},\ and\ \citenamefont {Carrera}(2006)}]{sanchez2006representative}%
  \BibitemOpen
  \bibfield  {author} {\bibinfo {author} {\bibfnamefont {X.}~\bibnamefont {Sanchez-Vila}}, \bibinfo {author} {\bibfnamefont {A.}~\bibnamefont {Guadagnini}},\ and\ \bibinfo {author} {\bibfnamefont {J.}~\bibnamefont {Carrera}},\ }\bibfield  {title} {\enquote {\bibinfo {title} {Representative hydraulic conductivities in saturated groundwater flow},}\ }\href@noop {} {\bibfield  {journal} {\bibinfo  {journal} {Reviews of Geophysics}\ }\textbf {\bibinfo {volume} {44}} (\bibinfo {year} {2006})}\BibitemShut {NoStop}%
\bibitem [{\citenamefont {Dagan}, \citenamefont {Fiori},\ and\ \citenamefont {Jankovic}(2013)}]{dagan2013}%
  \BibitemOpen
  \bibfield  {author} {\bibinfo {author} {\bibfnamefont {G.}~\bibnamefont {Dagan}}, \bibinfo {author} {\bibfnamefont {A.}~\bibnamefont {Fiori}},\ and\ \bibinfo {author} {\bibfnamefont {I.}~\bibnamefont {Jankovic}},\ }\bibfield  {title} {\enquote {\bibinfo {title} {Upscaling of flow in heterogeneous porous formations: Critical examination and issues of principle},}\ }\href {https://doi.org/https://doi.org/10.1016/j.advwatres.2011.12.017} {\bibfield  {journal} {\bibinfo  {journal} {Advances in Water Resources}\ }\textbf {\bibinfo {volume} {51}},\ \bibinfo {pages} {67--85} (\bibinfo {year} {2013})},\ \bibinfo {note} {35th Year Anniversary Issue}\BibitemShut {NoStop}%
\bibitem [{\citenamefont {Noetinger}(2013)}]{noetinger2013explicit}%
  \BibitemOpen
  \bibfield  {author} {\bibinfo {author} {\bibfnamefont {B.}~\bibnamefont {Noetinger}},\ }\bibfield  {title} {\enquote {\bibinfo {title} {An explicit formula for computing the sensitivity of the effective conductivity of heterogeneous composite materials to local inclusion transport properties and geometry},}\ }\href@noop {} {\bibfield  {journal} {\bibinfo  {journal} {Multiscale Modeling \& Simulation}\ }\textbf {\bibinfo {volume} {11}},\ \bibinfo {pages} {907--924} (\bibinfo {year} {2013})}\BibitemShut {NoStop}%
\bibitem [{\citenamefont {Klein}\ and\ \citenamefont {Randic}(1993)}]{Klein1993ResistanceD}%
  \BibitemOpen
  \bibfield  {author} {\bibinfo {author} {\bibfnamefont {D.~J.}\ \bibnamefont {Klein}}\ and\ \bibinfo {author} {\bibfnamefont {M.}~\bibnamefont {Randic}},\ }\bibfield  {title} {\enquote {\bibinfo {title} {Resistance distance},}\ }\href {https://api.semanticscholar.org/CorpusID:16382100} {\bibfield  {journal} {\bibinfo  {journal} {Journal of Mathematical Chemistry}\ }\textbf {\bibinfo {volume} {12}},\ \bibinfo {pages} {81--95} (\bibinfo {year} {1993})}\BibitemShut {NoStop}%
\bibitem [{\citenamefont {Spielman}\ and\ \citenamefont {Srivastava}(2008)}]{spielman2008graph}%
  \BibitemOpen
  \bibfield  {author} {\bibinfo {author} {\bibfnamefont {D.~A.}\ \bibnamefont {Spielman}}\ and\ \bibinfo {author} {\bibfnamefont {N.}~\bibnamefont {Srivastava}},\ }\bibfield  {title} {\enquote {\bibinfo {title} {Graph sparsification by effective resistances},}\ }in\ \href@noop {} {\emph {\bibinfo {booktitle} {Proceedings of the fortieth annual ACM symposium on Theory of computing}}}\ (\bibinfo {year} {2008})\ pp.\ \bibinfo {pages} {563--568}\BibitemShut {NoStop}%
\bibitem [{\citenamefont {Stauffer}\ and\ \citenamefont {Aharony}(2014)}]{stauffer2014introduction}%
  \BibitemOpen
  \bibfield  {author} {\bibinfo {author} {\bibfnamefont {D.}~\bibnamefont {Stauffer}}\ and\ \bibinfo {author} {\bibfnamefont {A.}~\bibnamefont {Aharony}},\ }\href@noop {} {\emph {\bibinfo {title} {Introduction to percolation theory: revised second edition}}}\ (\bibinfo  {publisher} {CRC press},\ \bibinfo {year} {2014})\BibitemShut {NoStop}%
\bibitem [{\citenamefont {De~Wit}(1995)}]{de1995correlation}%
  \BibitemOpen
  \bibfield  {author} {\bibinfo {author} {\bibfnamefont {A.}~\bibnamefont {De~Wit}},\ }\bibfield  {title} {\enquote {\bibinfo {title} {Correlation structure dependence of the effective permeability of heterogeneous porous media},}\ }\href@noop {} {\bibfield  {journal} {\bibinfo  {journal} {Physics of fluids}\ }\textbf {\bibinfo {volume} {7}},\ \bibinfo {pages} {2553--2562} (\bibinfo {year} {1995})}\BibitemShut {NoStop}%
\bibitem [{\citenamefont {Romeu}\ and\ \citenamefont {Noetinger}(1995)}]{romeu1995calculation}%
  \BibitemOpen
  \bibfield  {author} {\bibinfo {author} {\bibfnamefont {R.}~\bibnamefont {Romeu}}\ and\ \bibinfo {author} {\bibfnamefont {B.}~\bibnamefont {Noetinger}},\ }\bibfield  {title} {\enquote {\bibinfo {title} {Calculation of internodal transmissivities in finite difference models of flow in heterogeneous porous media},}\ }\href@noop {} {\bibfield  {journal} {\bibinfo  {journal} {Water Resources Research}\ }\textbf {\bibinfo {volume} {31}},\ \bibinfo {pages} {943--959} (\bibinfo {year} {1995})}\BibitemShut {NoStop}%
\bibitem [{\citenamefont {Wang}\ \emph {et~al.}(2017)\citenamefont {Wang}, \citenamefont {Wang}, \citenamefont {Liu}, \citenamefont {Wang}, \citenamefont {Wang},\ and\ \citenamefont {Cao}}]{wang2017finite}%
  \BibitemOpen
  \bibfield  {author} {\bibinfo {author} {\bibfnamefont {M.}~\bibnamefont {Wang}}, \bibinfo {author} {\bibfnamefont {Y.-F.}\ \bibnamefont {Wang}}, \bibinfo {author} {\bibfnamefont {Z.-F.}\ \bibnamefont {Liu}}, \bibinfo {author} {\bibfnamefont {X.-H.}\ \bibnamefont {Wang}}, \bibinfo {author} {\bibfnamefont {Y.}~\bibnamefont {Wang}},\ and\ \bibinfo {author} {\bibfnamefont {W.-D.}\ \bibnamefont {Cao}},\ }\bibfield  {title} {\enquote {\bibinfo {title} {Finite analytic numerical method for three-dimensional quasi-laplace equation with conductivity in tensor form},}\ }\href@noop {} {\bibfield  {journal} {\bibinfo  {journal} {Numerical Methods for Partial Differential Equations}\ }\textbf {\bibinfo {volume} {33}},\ \bibinfo {pages} {1475--1492} (\bibinfo {year} {2017})}\BibitemShut {NoStop}%
\bibitem [{\citenamefont {Broutin}\ \emph {et~al.}(2014)\citenamefont {Broutin}, \citenamefont {Devroye}, \citenamefont {Fraiman},\ and\ \citenamefont {Lugosi}}]{Broutin2014}%
  \BibitemOpen
  \bibfield  {author} {\bibinfo {author} {\bibfnamefont {N.}~\bibnamefont {Broutin}}, \bibinfo {author} {\bibfnamefont {L.}~\bibnamefont {Devroye}}, \bibinfo {author} {\bibfnamefont {N.}~\bibnamefont {Fraiman}},\ and\ \bibinfo {author} {\bibfnamefont {G.}~\bibnamefont {Lugosi}},\ }\bibfield  {title} {\enquote {\bibinfo {title} {Connectivity threshold of bluetooth graphs},}\ }\href {https://doi.org/10.1002/rsa.20459} {\bibfield  {journal} {\bibinfo  {journal} {Random Structures and Algorithms}\ }\textbf {\bibinfo {volume} {44}},\ \bibinfo {pages} {45 – 66} (\bibinfo {year} {2014})}\BibitemShut {NoStop}%
\bibitem [{\citenamefont {Knackstedt}, \citenamefont {Sahimi},\ and\ \citenamefont {Sheppard}(2000)}]{Knackstedt2000}%
  \BibitemOpen
  \bibfield  {author} {\bibinfo {author} {\bibfnamefont {M.~A.}\ \bibnamefont {Knackstedt}}, \bibinfo {author} {\bibfnamefont {M.}~\bibnamefont {Sahimi}},\ and\ \bibinfo {author} {\bibfnamefont {A.~P.}\ \bibnamefont {Sheppard}},\ }\bibfield  {title} {\enquote {\bibinfo {title} {Invasion percolation with long-range correlations: First-order phase transition and nonuniversal scaling properties},}\ }\href {https://doi.org/10.1103/PhysRevE.61.4920} {\bibfield  {journal} {\bibinfo  {journal} {Phys. Rev. E}\ }\textbf {\bibinfo {volume} {61}},\ \bibinfo {pages} {4920--4934} (\bibinfo {year} {2000})}\BibitemShut {NoStop}%
\bibitem [{\citenamefont {Seiffe}\ \emph {et~al.}(2022)\citenamefont {Seiffe}, \citenamefont {Ramírez}, \citenamefont {Sempé},\ and\ \citenamefont {Depino}}]{Seiffe2022}%
  \BibitemOpen
  \bibfield  {author} {\bibinfo {author} {\bibfnamefont {A.}~\bibnamefont {Seiffe}}, \bibinfo {author} {\bibfnamefont {M.~F.}\ \bibnamefont {Ramírez}}, \bibinfo {author} {\bibfnamefont {L.}~\bibnamefont {Sempé}},\ and\ \bibinfo {author} {\bibfnamefont {A.~M.}\ \bibnamefont {Depino}},\ }\bibfield  {title} {\enquote {\bibinfo {title} {Juvenile handling rescues autism-related effects of prenatal exposure to valproic acid},}\ }\href {https://doi.org/10.1038/s41598-022-11269-7} {\bibfield  {journal} {\bibinfo  {journal} {Scientific Reports}\ }\textbf {\bibinfo {volume} {12}} (\bibinfo {year} {2022}),\ 10.1038/s41598-022-11269-7}\BibitemShut {NoStop}%
\bibitem [{\citenamefont {Boschan}\ and\ \citenamefont {Noetinger}(2012)}]{boschan2012scale}%
  \BibitemOpen
  \bibfield  {author} {\bibinfo {author} {\bibfnamefont {A.}~\bibnamefont {Boschan}}\ and\ \bibinfo {author} {\bibfnamefont {B.}~\bibnamefont {Noetinger}},\ }\bibfield  {title} {\enquote {\bibinfo {title} {Scale dependence of effective hydraulic conductivity distributions in 3d heterogeneous media: a numerical study},}\ }\href@noop {} {\bibfield  {journal} {\bibinfo  {journal} {Transport in porous media}\ }\textbf {\bibinfo {volume} {94}},\ \bibinfo {pages} {101--121} (\bibinfo {year} {2012})}\BibitemShut {NoStop}%
\end{thebibliography}%

\end{document}